\documentclass[pre,amsfonts,english,showpacs,reprint,floatfix]{revtex4-1}
\usepackage[T1]{fontenc}
\usepackage[latin1]{inputenc}
\usepackage{amsmath}
\usepackage{graphicx}
\usepackage{amssymb}
\usepackage{verbatim}

\usepackage{soul}
\makeatletter
\usepackage{babel}
\usepackage[normalem]{ulem} \makeatother

\newcommand{\eref}[1]{Eq.\ (\ref{#1})}
\newcommand{\Eref}[1]{Equation (\ref{#1})}
\newcommand{\fref}[1]{Fig.\ \ref{#1}}
\newcommand{\Fref}[1]{Figure \ \ref{#1}}
\newcommand{\sref}[1]{Sec. \ \ref{#1}}
\newcommand{\aref}[1]{Appendix \ \ref{#1}}
\newcommand{\eqs}{Eqs.\ }

\newcommand{\fl}{}
\newcommand{\rs}{r}
\newcommand{\rss}{r^2}

\makeatletter


\usepackage{babel}
\makeatother
\begin{document}

\title[]{Non-equilibrium ensemble inequivalence and density large deviations in the ABC model}

\author{O Cohen and D Mukamel}

\address{Department of Physics of Complex Systems, Weizmann Institute of Science, Rehovot 7610001, Israel}

\begin{abstract}
We consider the one-dimensional driven ABC model under particle-conserving and particle-non-conserving
 processes.
 Two limiting cases are studied: (a) the rates of the non-conserving processes are vanishingly slow compared with the conserving processes in the thermodynamic
 limit and (b) the two rates are comparable. For case (a) we provide a detailed analysis of the phase diagram and the large deviations function of the
 overall density,  $\mathcal{G}(r)$.
  The phase diagram of the non-conserving model, derived from $\mathcal{G}(r)$, is found to be different from the conserving one. This difference, which stems
 from the non-convexity of $\mathcal{G}(r)$, is analogous to ensemble inequivalence found in equilibrium systems with
 long-range interactions. An outline of the analysis of case (a) was given in an earlier letter.
 For case (b) we show that unlike the conserving model, the non-conserving model exhibits a moving density profile in the steady-state with a velocity that
 remains finite in the thermodynamic limit. Moreover, in contrast with case (a), the critical lines of the conserving and non-conserving models do not coincide.
 These are new features which are present only when the rates of the conserving and non-conserving processes are comparable.
 In addition, we analyze $\mathcal{G}(r)$ in case (b)  using macroscopic fluctuations theory.
  Much of the derivation presented in this paper is applicable to any driven-diffusive system coupled to an external particle-bath via a slow dynamics.

\end{abstract}


\maketitle

\section{Introduction}
Systems that are driven out of equilibrium by an external field, such
as temperature gradient or electric field, exhibit in many cases long-range correlations in their steady-states,
 even when their dynamics is strictly local.
 This has been demonstrated in numerous studies both for specific models and in more generic settings \cite{Spohn1983,domb1995statistical,lepri2003thermal,derrida2007non,blythe2007nonequilibrium,dhar2008heat,Sadhu2011}.
 Studying generic steady-state features that result from these long-range correlations would be of great interest.
 A useful insight into this behaviour may be gained by comparing it to that of equilibrium system,
 where long-range correlations appear either in systems with short-range interactions at criticality or
 in systems with explicit long-range interactions. The latter type bears more similarity to {\it driven systems}, as in both cases
 the long-range correlations appear generically, i.e. also away from any phase transition points.

Equilibrium systems with long-range interactions are those where the two body potential decays at large distance, $R$, as $1/R^{d+\sigma}$, with
$-d\leq\sigma\leq0$ in $d$ dimensions. One consequence of this long-range decay is non-additivity, whereby
the energy does not increase linearly with the system's size.
Non-additivity may cause various ensembles of the same long-range interacting system to exhibit different phase diagrams, as has been shown in numerous studies
\cite{Antonov1962,LyndenBell1968,Thirring1970,Hertel1971,LyndenBell1999,
Thirring2003,Barre2001,Barre2002,Mukamel2005,Ellis2004,Touchette2004,Touchette2005,Bouchet2005,Mukamel2008}.
This occurs, for instance, when the microcanonical entropy is not a convex function of the energy for a certain range of energies,
leading to negative specific heat in the microcanonical ensemble.
This is in contrast with the canonical ensemble, where the entropy is inherently convex and the specific heat is non-negative.
 Similar effects can be found when comparing the canonical and the grand-canonical ensembles
\cite{Misawa2006,Grosskinsky2008,Lederhendler2010a}.

In a recent letter \cite{Cohen2012} we have demonstrated that a phenomenon similar to ensemble inequivalence exists in
 a specific driven system, known as the ABC model \cite{Evans1998,Evans1998b}.
The ABC model is defined on a one-dimensional lattice of length $L$, where each site is occupied by one particle of type $A$, $B$ or $C$.
The model evolves by sequential updates whereby particles on randomly chosen neighbouring sites are exchanged with the following rates:
\begin{equation}
\label{eq:ABCdynamics}
AB\overset{q}{\underset{1}{\rightleftarrows}}BA\,\qquad\,
BC\overset{q}{\underset{1}{\rightleftarrows}}CB\,\qquad\,
CA\overset{q}{\underset{1}{\rightleftarrows}}AC.
\end{equation}
The ABC model is often studied in the limit of weak-asymmetry where the asymmetry in the rates
scales with the size of the system, $L$, as $q=\exp\left(-\beta/L\right)$ \cite{Clincy2003}.
With this scaling the model undergoes a transition between
a phase where the particles are homogenously distributed in the system and a phase
where the three species phase separate into three macroscopic domains \cite{Clincy2003}. The value of $\beta$ where the transition occurs is a function
of the average densities of particles, defined as $r_\alpha\equiv N_\alpha/L$ for $\alpha=A,B,C$, where $N_\alpha$ is the overall
number of particles of type $\alpha$.
A unique property of the ABC model is that when the number of particles of the three species are equal, $N_A=N_B=N_C$,
the model obeys detailed balance with respect to an effective Hamiltonian with long-range interactions.
 This special equilibrium point provides an analytical framework for investigating the mechanism behind
 long-range correlations in driven systems.

The ABC model has been generalized to include particle non-conserving process in \cite{Lederhendler2010a,Lederhendler2010b}.
In this generalized model sites can also be occupied by inert vacancies, denoted by $0$, whose dynamics is defined as
\begin{equation}
\label{eq:vacancyexchange}
A0\overset{1}{\underset{1}{\rightleftarrows}}0A,
\qquad B0\overset{1}{\underset{1}{\rightleftarrows}}0B,
\qquad C0\overset{1}{\underset{1}{\rightleftarrows}}0C.
\end{equation}
In addition, triplets of particles can evaporate and condense with the following rates:
\begin{equation}
ABC\overset{p \, e^{-3\beta \mu}}{\underset{p}{\rightleftarrows}}000,\label{eq:addremoverates}
\end{equation}
where $p$ is a rate parameter and $\mu$ plays the role of the chemical potential.
Studies of this model in the {\it equal-densities} regime  revealed inequivalence between the phase diagram of a {\it conserving model},
  defined by rules (\ref{eq:ABCdynamics})-(\ref{eq:vacancyexchange}), and that of a {\it non-conserving model}, defined
 by rules (\ref{eq:ABCdynamics})-(\ref{eq:addremoverates}).
 The two models correspond to the canonical and grand-canonical ensembles of the ABC Hamiltonian, respectively.
 The generalized ABC model has then been analyzed for non-equal densities, where an effective Hamiltonian cannot be defined \cite{Cohen2012},
 and was demonstrated to exhibit a similar ensemble inequivalence as in the equal-densities case.

In this paper we generalize the analysis of \cite{Cohen2012} and present a detailed study of the ABC model with particles non-conserving processes.
In the first part of the paper we study the non-conserving ABC model
 in the limit where the ratio between the time-scale of the non-conserving dynamics, given by $1/p$,
and that of the conserving dynamics, given by $\tau \sim L^2$, vanishes in the thermodynamic limit. We thus consider $p\sim L^{-\gamma}$ with $\gamma>2$.
 In this limit, we are able to compute the large deviations function (rate function) of the overall density of particles, $\mathcal{G}(r)$ with $r \equiv (N_A+N_B+N_C)/L$, and
 derive from it the exact phase diagram of the non-conserving model.
 This is done by analyzing the dynamics of $r$, which can be effectively represented as a one-dimensional random walk in
 a potential. The form of the potential is derived from the coarse-grained density profile of the ABC model, computed in \cite{Cohen2011a}.
 The non-conserving phase diagram for non-equal densities, derived from $\mathcal{G}(r)$, is similar to that obtained for equal densities, excluding several new features discussed below.
As in the equal-densities case and similarly to equilibrium systems with long-range interactions,
the conserving and the non-conserving models undergo the same second order transition and become inequivalent when the transition in the non-conserving model
turns into first order.
This suggests that the phenomenon of ensemble inequivalence, which characterizes
 many long-range interacting systems, may be found in other driven diffusive systems that exhibit long-range correlations. %A brief account of this study is given in \cite{Cohen2012}.

In the second part of the paper we study the non-conserving ABC model in the limit where the rates of the conserving and non-conserving dynamics are comparable,
namely for $p=\phi L^{-2}$ with $\phi$ being an arbitrary parameter.
In this case the conserving and the non-conserving models exhibit different critical lines and steady-state density profiles.
Such inequivalence is expected to be found only in non-equilibrium systems.
Interestingly, the density profile of the non-conserving model is found to exhibit a drift velocity in its ordered phase.
The velocity remains finite even in the thermodynamic limit and can be computed exactly along the critical line.
This is in contrast with the conserving ABC model, where for $r=1$ the drift velocity has been shown to vanish as $1/L$ \cite{Bodineau2011} and has been
excluded in the thermodynamic limit \cite{Bertini2012}. This can be shown to be valid also for the conserving model with $r<1$ using the mapping between the $r=1$ and $r<1$ cases, discussed below.

The probability of a rare number of particles in the non-conserving model, $P(r)\sim e^{L\mathcal{G}(r)}$, can be studied for $p=\phi L^{-2}$ using the macroscopic fluctuation theory
 \cite{freuidlin1994random,bertini2001fluctuations,jordan2004transport,bertini2005current,tailleur2008mapping,touchette2012large}.
By analyzing the instanton path leading to a rare value of $r$,
 we derive an expansion of $\mathcal{G}(r)$ in powers of $\phi$ for $\phi\ll 1$. The lowest order in this expansion is identical to the expression obtained
for $p\sim L^{-\gamma}$ and $\gamma>2$.
This fact serves as a proof that $\gamma>2$ is indeed the limit where the conserving and nonconserving time-scales are
well separated.
  In the homogenous phase the corrections to $\mathcal{G}(r)$ vanish for all $\phi$, yielding an the same expression for $\gamma>2$ and $\gamma=2$.
This conclusion is expected to remain valid for a large class of driven diffusive systems that exhibit a homogenous phase.

This paper is organized as follows. We first provide a brief review of the ABC model and
previous studies of the equal-densities regime in \sref{sec:ABC}. We study the phase
diagram of the non-conserving ABC model for non-equal densities and in the limit of very slow non-conserving
dynamics ($\gamma>2$) in \sref{sec:slow}.
In \sref{sec:fast_evaportation} we study the model in the limit where the conserving and the non-conserving dynamics
occur on comparable time-scales ($\gamma=2$). Concluding remarks and outlook are given in \sref{sec:conc}.

\section{The ABC model with equal densities}
\label{sec:ABC}
 The standard ABC model is defined by the dynamical rules in \eref{eq:ABCdynamics}.
For $q=1$, this dynamics yields a
homogeneous steady state, in which the particles are distributed uniformly on the lattice.
On the other hand, for $q \ne 1$ the model relaxes in the $L\to \infty$ limit to a state where the particles phase separate into three domains.
The domains are arranged clockwise for $q<1$, i.e. $AA\ldots ABB\ldots BCC\ldots C$, and
counterclockwise for $q>1$. Throughout this paper we assume $q<1$. The case
of $q>1$ is obtained by considering a system with a driving field $q'=1/q<1$ and
exchanging, say, the $B$ and $C$ labels.

As a result of the dynamical asymmetry,
 the model exhibits in general non-vanishing steady-state currents of particles.
The current of particles of a given species is proportional to the rate at which the particles of this species perform a full clockwise trip minus the rate of the counter-clockwise trip, yielding
\begin{equation}
J_{\alpha}\sim q^{N_{\alpha+1}}-q^{N_{\alpha+2}}, \label{eq:currents}
\end{equation}
where $\alpha$ runs cyclically over $A,B,C$ and $N_\alpha$ denotes the overall number of particles of species $\alpha$.
In the {\it equal-densities case}, where $N_{A}=N_{B}=N_{C}=L/3$, the currents vanish and
the dynamics can be shown to obey detailed balance with respect to an effective
long-range Hamiltonian  given by
\begin{equation}
 \label{eq:trans_invariant_H}
\mathcal{H}\left( {\boldsymbol \zeta }
\right)=\sum_{i=1}^{L}\sum_{k=1}^{L-1}\frac{k}{L}
\left(A_{i}B_{i+k}+B_{i}C_{i+k}+C_{i}A_{i+k}\right).
\end{equation}
Here ${\boldsymbol \zeta }=\left\{{ \zeta}_i\right\}_{i=1}^{L}$ denotes a microstate of the system such that $\zeta_i=A,B$ or $C$, and
\begin{equation}
\label{eq:projection_op}
X_i= \left\{ \begin{array}{ccc}
1 &  & \zeta_{i}=X\\
0 &  & \zeta_{i}\neq X
\end{array}\right.  \qquad \mathrm{for} \; X=A,B,C.
\end{equation}
The steady-state probability measure is
given in terms of the Hamiltonian by  $P({\boldsymbol \zeta })\propto q^{\mathcal{H}( {\boldsymbol \zeta })}$ \cite{Evans1998}.

\begin{figure}
\noindent
\begin{centering}\includegraphics[scale=0.6]{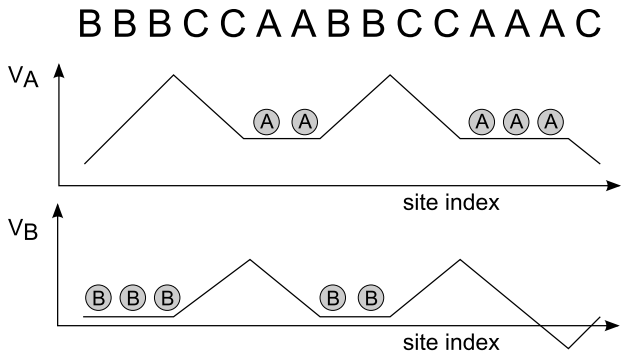}\par\end{centering}
\caption{
\label{fig:potential_ABC} Effective potential exerted on the $A$ particles, $V_{A,i}=\sum_{k=1}^{i-1} B_{k}-C_{k}$,
and on the $B$ particles,  $V_{B,i}=\sum_{k=1}^{i-1} C_{k}-A_{k}$, drawn for a specific configuration of a system of
size $L=15$. The configuration itself is written above the two figures.}
\end{figure}

The Hamiltonian in \eref{eq:trans_invariant_H} can also be written in a more instructive form as
\begin{eqnarray}
 \label{eq:H1}
\mathcal{H}\left( {\boldsymbol \zeta } \right) &= \frac{1}{2} \sum_{i=1}^{L-1}\sum_{k=1}^{i-1} \big[ & A_{i}(B_{k}-C_{k}) +B_{i}(C_{k}-A_{k}) \nonumber \\
&&+C_{i}(A_{k}-B_{k})\big],
\end{eqnarray}
where every particle appears to be positioned in a potential well created by the particles of the two other species, as illustrated in \fref{fig:potential_ABC} for a specific configuration.
The factor $1/2$ corrects for the double-counting of the interactions. Clearly, this potential picture breaks down for non-equal densities, where the system relaxes to a non-equilibrium steady-state.

We consider in this paper the limit of weak asymmetry, where $q\to 1$ in the $L\to \infty$ limit
as $q=\exp\left(-\beta/L\right)$ with $\beta$ being a positive parameter \cite{Clincy2003}.
As a result of this scaling, the energy and the entropy of the system become comparable in size in the thermodynamic limit.
The parameter $\beta$ which sets the ratio between the energy and entropy can be regarded as the inverse temperature of the system.
This rescaling is similar to the
Kac's prescription often employed systems with long-range interactions \cite{Kac1963}.
 For equal densities, the model exhibits in the $L\to \infty$ limit a second order phase transition
at $\beta=\beta_c=2\pi\sqrt{3}$ between a homogeneous state, where entropy dominates, and an ordered state
which is dominated by the energy term \cite{Clincy2003}.

Ensemble inequivalence can be studied in the ABC model by comparing its canonical and grand-canonical phase diagrams.
To this end we consider a generalization of the original model
 where sites may also be vacant \cite{Lederhendler2010a}.
In this case,  particles can hop into and out of vacant sites, denoted by $0$, with symmetric rates given by \eref{eq:vacancyexchange}.
The number of vacancies is $N_0 \equiv L-N$, where $N \equiv N_A+N_B+N_C\leq L$.
We refer to this model, which consists of rules (\ref{eq:ABCdynamics}) and (\ref{eq:vacancyexchange}), as the {\it conserving ABC model}.

 The steady-state properties of the conserving model can be derived by mapping it onto the standard ABC model in which $N=L$.
 This is a many-to-one mapping which consists of removing the vacancies in a micro-configuration, $\boldsymbol \zeta$, of the generalized model of length $L$, yielding a configuration
 of length $N$, denoted by ${\bf f}( \boldsymbol \zeta ) $.
Using the fact  that the vacancies evolve by a simple diffusion (\ref{eq:vacancyexchange})
 and are thus homogenously distributed in the steady-state, it can easily be shown that the steady-state measure of the generalized model is given by
\begin{equation}
\label{eq:mapping}
P(\boldsymbol \zeta;N_0=L-N) =  P( {\bf f}( \boldsymbol \zeta );N_0=0)   /  {L \choose N},
\end{equation}
where $P( \boldsymbol \zeta ;N_0=0)$ denotes the steady-state measure of the standard ABC model of size $N$. For equal-densities the latter is given by $P( \boldsymbol \zeta ;N_0=0)\propto q^{\mathcal{H}( {\boldsymbol \zeta })}$.
By defining an effective inverse temperature  $\beta'$ as $q=\exp(\beta/L)=\exp(\beta r/N)\equiv \exp(\beta'/N)$, where $r=N/L$, we conclude that the $N$-size
system has an effective inverse temperature  $\beta'=\beta r$. Similarly the average
densities of the $N$-size system are $r_\alpha'=r_\alpha/r$ for
$\alpha=A,B,C$.
Using this mapping the critical point of the equal-densities standard ABC model \cite{Clincy2003}, $\beta_c=2\pi\sqrt{3}$, is mapped onto a critical line in the equal-densities conserving model, given by
\begin{equation}
\label{eq:bc_equal}
\beta_c=2\pi\sqrt{3}/r.
\end{equation}

\begin{figure}[t]
\noindent
\begin{centering}\includegraphics[scale=0.6]{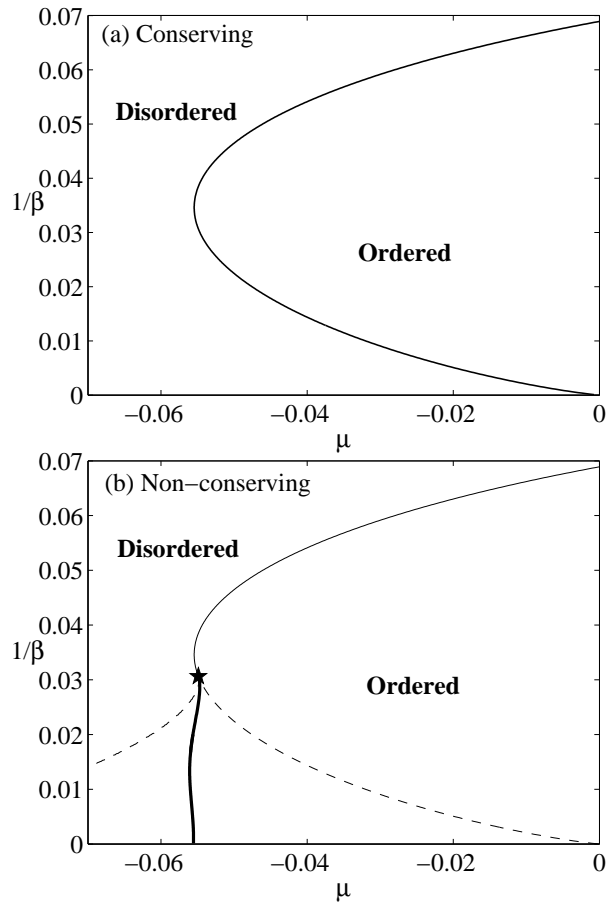}\par\end{centering}
\caption{Phase diagram of the conserving (a) and non-conserving (b) ABC models with equal densities, $N_A=N_B=N_C$, plotted in the ($\mu,1/\beta$) plane.
The two diagrams show the same second order transition line (thin solid line).
In the non-conserving model the transition line turns into a first order (thick solid line) at a tricritical point ($\star$). The boundaries of the
coexistence region, where both the homogenous and ordered profiles are locally stable, are denoted in the non-conserving model by dashed lines.
\label{fig:MuT_equal}}
\end{figure}

The {\it non-conserving ABC model} is defined by allowing for
evaporation and deposition of particles, performed in triplets of neighboring particles according to \eref{eq:addremoverates}.
For $N_A=N_B=N_C$, this specific type of non-conserving process can be shown to maintain detailed balance
with respect to the Hamiltonian
\begin{equation}
\mathcal{H}_{GC}\left( {\boldsymbol \zeta } \right) = \mathcal{H}\left( {\boldsymbol \zeta } \right)-\mu N L,
\end{equation}
where $\mathcal{H}\left( {\boldsymbol \zeta } \right)$ is defined in \eref{eq:H1} \footnote{ In \cite{Lederhendler2010b} the grand-canonical Hamiltonian was defined based on the
Hamiltonian in \eref{eq:trans_invariant_H}. In this case one has to add a term of the form $-\frac{1}{6}N(N-1)$ from $\mathcal{H}_{GC}\left( {\boldsymbol \zeta } \right)$
in order to allow for local dynamics whose rate does not depend on $N$. This additional term is not necessary when using the Hamiltonian in \eref{eq:H1}, since the potential
landscape changes only locally when adding or removing a triplet of $ABC$ particles. This suggests the \eref{eq:H1} is a more natural representation of the stationary measure
of the equal-densities ABC model.}
. This allows
 one to study the non-conserving model using equilibrium techniques.
It has been shown in \cite{Lederhendler2010a} that for equal-densities the non-conserving model also exhibits a second order transition line at $\beta_c=2\pi\sqrt{3}/r$
for $r\geq r_{TCP}$, where $r_{TCP}=1/3$ is the tricritical point.
Below this point, for $r<r_{TCP}$, the non-conserving model exhibits
a first order transition, whereas the transition in the conserving model remains second order.
The resulting phase diagrams of the conserving and non-conserving models, which have been derived for equal densities in \cite{Lederhendler2010b}, are plotted in \fref{fig:MuT_equal}.
The generalization of this phase diagram to arbitrary densities, where one cannot use the Hamiltonian in \eref{eq:trans_invariant_H}, is discussed in the next section.

In order to compare the two ensembles, the canonical phase diagram is plotted in \fref{fig:MuT_equal}  as a function of the chemical potential, which is defined as
\begin{equation}
\mu(r)=d\mathcal{F}(r)/dr,
\end{equation}
where $\mathcal{F}(r)$ is free energy of the conserving ABC model.
For equal densities, the latter has been computed in \cite{Lederhendler2010b} in the $L\gg1$ limit
 based on a continuum description of the model as
 \begin{eqnarray}
\fl &&\mathcal{F}(r) =    E[ \boldsymbol \rho^\star(x,r)]  - \frac{1}{\beta} S[ \boldsymbol \rho^\star(x,r)  ] . \nonumber
\end{eqnarray}
Here $\boldsymbol \rho^\star (x,r) \equiv (\rho_A^\star(x,r),\rho_B^\star(x.r),\rho_C^\star(x,r))$ is the steady-state profile of the conserving model with particle density $r$, whose
form is discussed in \sref{sec:fol}.  The functionals $S$ and $E$ correspond to the entropy and the energy per particle for a given density profile, respectively.
The former is derived from combinatorial considerations, yielding  $S[ \boldsymbol \rho]=-\sum_\alpha\int_0^1 dx \rho_\alpha(x) \log\rho_\alpha(x) - \int_0^1 dx \rho_0 (x) \log \rho_0 (x)$, and
the latter is derived from the Hamiltonian in \eref{eq:H1}, yielding $E[\boldsymbol \rho]=\frac{1}{2}\sum_\alpha\int_0^1 dx \int_0^x dy \rho_\alpha(x) [\rho_{\alpha+1}(y)-\rho_{\alpha+2}(y)]$.
Here and below $\alpha$ runs cyclically over $A,B$ and $C$ and $\rho_0 (x)\equiv 1-\sum_\alpha \rho_\alpha(x)$.

\section{ABC model with non-equal densities and vanishingly slow non-conserving dynamics ($p\sim L^{-\gamma}, \, \gamma >2$)}
\label{sec:slow}

In this section we study the steady-state properties of the ABC model for arbitrary densities both under conserving dynamics and non-conserving dynamics
 whose rate, $p$, is vanishingly slower than the diffusive dynamics for $L\to \infty$.

For arbitrary densities, in absence of a Hamiltonian which describes the steady-state measure of the model,
the analysis of the model is based on a continuum description which is valid in general for diffusive systems such as the weakly-asymmetric ABC model.
In the continuum limit the probability of a micro-configuration $\boldsymbol \zeta(t)$ is approximated by a smooth density profile defined as
\begin{equation}
\label{eq:Peta}
Pr(\zeta_{i}(t)=\alpha)\simeq \rho_\alpha(i/L,t/L^2),
\end{equation}
for $\alpha=A,B,C$. The evolution of the coarse-grained density profile, $\rho_\alpha(x,\tau)$, is given in the conserving model by the following hydrodynamic equation \cite{Clincy2003}:
\begin{equation}
\label{eq:meanfieldA}
\partial_{\tau} \rho_\alpha = \beta \partial_x\left[\rho_\alpha\left(\rho_{\alpha+1}-\rho_{\alpha+2}\right)\right]+ \partial_x^2\rho_\alpha
\end{equation}
The density profile is periodic, $\rho_\alpha(x+1,\tau)=\rho_\alpha(x,\tau)$, and due to exclusion it obeys $\rho_0(x,\tau)=1-\sum_\alpha \rho_\alpha(x,\tau)\leq 1$.
\Eref{eq:meanfieldA} has been shown to be exact in the
 $L\to\infty$ limit for equal densities and $r=1$ \cite{Clincy2003,Ayyer2009}, and has been argued to remain valid even for arbitrary average densities \cite{Ayyer2009,Bertini2012}.
The mapping discussed in \sref{sec:ABC} between the $r=1$ and the $r<1$ cases implies that the steady-state properties resulting from the study of \eref{eq:meanfieldA} for $r=1$ can be mapped onto the $r<1$ case.

Within the hydrodynamic framework, the addition of the non-conserving dynamics in (\ref{eq:addremoverates}) yields
 \begin{eqnarray}
\label{eq:meanfieldB}
\partial_{\tau} \rho_\alpha &=& \beta \partial_x\left[\rho_\alpha\left(\rho_{\alpha+1}-\rho_{\alpha+2}\right)\right]+ \partial_x^2\rho_\alpha \\
 &&+ L^2 p \left(\rho_0^3-e^{-3\beta\mu}\rho_A\rho_B\rho_C\right), \nonumber
\end{eqnarray}
where the $L^2$ factor of the non-conserving term is due to the rescaling of time, $\tau=t/L^2$, and of space, $x=i/L$. Using the same reasoning discussed in \cite{Bertini2012} for the conserving model, the local-equilibrium approximation is
expected to be valid when the rate of the non-conserving dynamics is either slower or equal to the inverse of the diffusive time-scale,
i.e. for $p\sim L^{-\gamma}$ and $\gamma \geq 2$. \Eref{eq:meanfieldB} is thus assumed to be valid for $\gamma \geq 2$.

In the following subsections we analyze the phase diagram of the non-conserving model for $\gamma>2$, first in \sref{sec:expand} by expanding \eref{eq:meanfieldB} around its homogenous solution and
then in \sref{sec:fol} by computing $\mathcal{G}(r)$ from the steady-state solution of \eref{eq:meanfieldA}.

\subsection{Critical expansion of the ABC model}
\label{sec:expand}
\subsubsection{Conserving dynamics}
We consider first the conserving ABC model, described by \eref{eq:meanfieldA}. In order to find the
critical of the model point one can expand its density profile around the homogenous solution $\rho_\alpha(x) = r_\alpha$. Following the approach presented in \cite{cohen2012ensemble} for the equal-densities case,
we expand the profile in terms of Fourier modes in $x$-space and in terms of the eigenvectors of the matrix
 $\mathcal{T}_{\alpha,\alpha'} \equiv \delta_{\alpha+1,\alpha'}-\delta_{\alpha+2,\alpha'}$ in the species-space.
The density profile can thus be written as,
\begin{eqnarray}
\label{eq:rho_expand}
\fl \left(\begin{array}{c}
\rho_{A}\\
\rho_{B}\\
\rho_{C}
\end{array}\right)&=&\sum_{m=-\infty}^{\infty}e^{2\pi imx}\Big[\frac{a_{m}(\tau)}{\sqrt{3}}\left(\begin{array}{c}
1\\
e^{-2\pi i/3}\\
e^{2\pi i/3}
\end{array}\right) \\
&&+\frac{a_{-m}^{\star}(\tau)}{\sqrt{3}}\left(\begin{array}{c}
1\\
e^{2\pi i/3}\\
e^{-2\pi i/3}
\end{array}\right)+\frac{b_{m}(\tau)}{3}\left(\begin{array}{c}
1\\
1\\
1
\end{array}\right)\Big]. \nonumber
\end{eqnarray}
There are several points to note about this expansion. Integrating \eref{eq:rho_expand} over $x$, yields
\begin{equation}
\label{eq:b0_interp}
\left(\begin{array}{c}
r_{A}\\
r_{B}\\
r_{C}
\end{array}\right)=\frac{b_0}{3}
\left(\begin{array}{c}
1\\
1\\
1
\end{array}\right)
+\frac{2}{\sqrt{3}} |a_0|
\left(\begin{array}{c}
\cos \theta\\
\cos(\theta-\frac{2\pi}{3})\\
\cos(\theta+\frac{2\pi}{3})
\end{array}\right),
\end{equation}
where $\theta=\mathrm{arg}(a_0)$. Summing \eref{eq:b0_interp} over $\alpha=A,B,C$ yields $b_0=r_A+r_B+r_C=r$, which implies that $b_0$ represents the average
density.  On the other hand,
$a_0$ sets the deviation of $r_A,r_B$ and $r_C$ from the equal-densities point.
 Similarly for $m\neq 0$, the values of $b_m$ set the profile
of the vacancies while $a_m$ set the local deviation from equal densities.
One reason to choose the eigenvectors of $\mathcal{T}$ is because they
 decompose the deviations from equal densities in a symmetric way that does not prefer any specific species.

Inserting the above expansion (\ref{eq:rho_expand}) into \eref{eq:meanfieldA} we find that the evolution of $a_m$ and $b_m$ is decoupled and given by
\begin{eqnarray}
\label{eq:complex_am_full}
\partial_{\tau}a_{m}&=&-\frac{2\pi m}{3}\big[\left(6\pi m-\sqrt{3}\beta r\right)a_{m} \\
&&\qquad +3\beta\sum_{m'=-\infty}^{\infty}a_{-m-m'}^{\star}a_{m'}^{\star}\big],\nonumber \\
\label{eq:complex_bm}
\partial_{\tau} b_m &=& - 4\pi ^2 m^2 b_m.
\end{eqnarray}
The latter equation reflects the fact that the inert vacancies perform an unbiased diffusion and hence their
steady-state profile is flat, $\rho_0(x,\infty)=1-r$, or equivalently $b_m(\infty)=0$ for $m\neq0$.

In order to find the critical line one has to consider small perturbations around the homogenous profile by taking $a_m\ll 1$.
For non-equal densities, where $a_0\neq0$, the non-linear term in \eref{eq:complex_am_full} contributes to the lowest order equation of $a_m$.
To lowest order in $a_m$ one finds a linear dependence between $a_m$ and $a_{-m}^\star$ given by,
\begin{eqnarray}
\label{eq:complex_am}
\fl \qquad \left(\begin{array}{c}
\partial_{\tau}a_{m}\\
\partial_{\tau}a_{-m}^{\star}
\end{array}\right)= -\frac{2\pi m}{3} A_m \left(\begin{array}{c}
a_{m}\\
a_{-m}^{\star}
\end{array}\right)+\mathcal{O}(a_{m}^2). \nonumber
\end{eqnarray}
where
\begin{equation}
A_m \equiv \left(\begin{array}{cc}
6\pi m-\sqrt{3}\beta r & 6\beta a_{0}^{\star}\\
-6\beta a_{0} & 6\pi m+\sqrt{3}\beta r
\end{array}\right).
\end{equation}
The critical point of the model is obtained when the highest of the eigenvalues of $-\frac{2\pi m}{3}A_m$ vanishes and as a result the amplitude of the corresponding eigenvector becomes unstable.
Out of all the eigenvalues of $-\frac{2\pi m}{3} A_m$, given by
\begin{equation}
\epsilon_{\pm}^{(m)}= -\frac{2\pi m}{3} \left( 6 m \pi \mp  \beta \sqrt{3 \left( r^2-12|a_{0}|^{2}\right)}\right) .
\end{equation}
the highest one is $\epsilon_+^{(1)}$ and it vanishes at
\begin{equation}
\label{eq:bc_cons_non-equal}
\beta= \beta_c \equiv \frac{2\pi \sqrt{3}}{\sqrt{r^2-12|a_{0}|^{2}}} .
\end{equation}
\Eref{eq:bc_cons_non-equal} can be written in terms of $r_\alpha$ by  taking the square of \eref{eq:b0_interp}, yielding
\begin{equation}
\left|a_0\right|^2=\frac{1}{2}\sum_{\alpha=A,B,C}(r_\alpha-r/3)^2.
\end{equation}
As expected, the critical line in \eref{eq:bc_cons_non-equal} coincides with the result for $r=1$ obtained in \cite{Clincy2003}.

For $\beta=\beta_c(1+\delta)$ with $\delta \ll 1$, the existence of an ordered profile whose amplitude vanishes with $\delta$ corresponds to a continuous transition at $\beta=\beta_c$.
To leading order in $\delta$, only the eigenvector corresponding to $\epsilon_+^{(1)}$ is excited and the other Fourier modes are driven by it through the nonlinear terms in  \eref{eq:complex_am_full}.
Denoting the amplitude of this eigenvector by $\varphi$, one can expand \eref{eq:complex_am_full} in powers of $\varphi$, yielding
\begin{equation}
\partial_{\tau} \varphi =  4\pi^2 \delta \varphi + G^{c}_4(r,a_0) \varphi \left| \varphi\right|^2 + G^{c}_6(r,a_0) \varphi \left| \varphi\right|^4 + O(\varphi^7),
\end{equation}
where the superscript $c$ refers to the conserving dynamics. In \aref{sec:high_order}, the form of $G^{c}_i(r,a_0)$ is obtained explicitly for $i=4$ and studied numerically for $i=6$.
The case where $G^{c}_4(r,a_0)>0$ yields an ordered profile whose amplitude vanishes as $\delta^{1/2}$, whereas for $G^{c}_4(r,a_0)<0$ the steady-state
profile has a finite amplitude. One can thus identify the point where $G^{c}_4(r,a_0)=0$ and $\beta=\beta_c$ as the tricrtical point of the model,
 below which, for $G^{c}_4(r,a_0)< 0$, the critical line is preempted by a first order transition. The
condition of $ \beta=\beta_c$ and $G^{c}_4(r,a_0)<0$ is shown in \aref{sec:high_order} to correspond to
\begin{equation}
\label{eq:tcl_cons}
2(r_A^3+r_B^3+r_C^3)>(r_A+r_B+r_C)(r_A^2+r_B^2+r_C^2).
\end{equation}
This condition coincides with that obtained for $r=1$ in \cite{Clincy2003}.

\begin{figure}[t]
\noindent
\begin{centering}\includegraphics[scale=0.6]{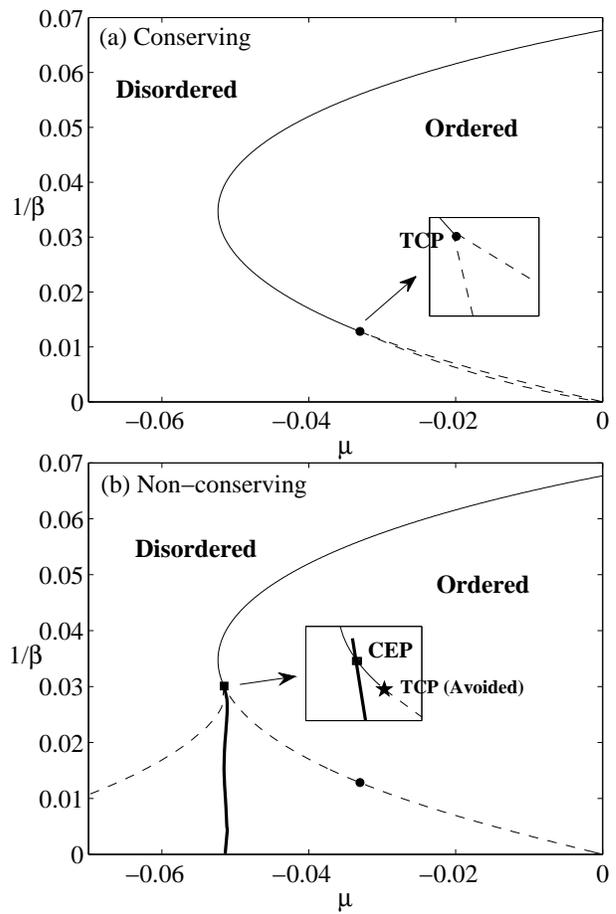}\par\end{centering}
\caption{\label{fig:MuT_non-equal} Phase diagram of the conserving (a) and non-conserving (b) ABC models for in two-equal-densities case,
 $r_A = r_B = r/3-0.01$ and $r_C = r/3+0.02$ or equivalently $a_0=0.01 \sqrt{3} e^{i\pi /3}$ , plotted in the ($\mu,1/\beta$) plane.
The two diagrams show the same second order transition line (thin solid line).
In the conserving model the transition line turns at the tricritical point ($\bullet$) into a first order transition line, which cannot be computed without knowing the full large deviations function of the ABC model.
In the non-conserving model the transition line turns into a first order (thick solid line) at a critical-end-point ($\blacksquare$), which appears at higher temperature than the
tricritical point ($\star$). Both diagrams are supplemented by an illustration
of the region where the first and second order transition lines meet. The
 boundaries of the coexistence regions, where both the homogenous and the ordered profiles are locally stable, are denoted by dashed lines.
}
\end{figure}

The critical line in \eref{eq:bc_cons_non-equal} and the above tricritical point, where $\beta=\beta_c$ and $G^{c}_4(r,a_0)=0$,
 are plotted for a specific value of $a_0$ in \fref{fig:MuT_non-equal}a. For comparison with the non-conserving model
the conserving phase diagram is plotted as a function of the parameter $\mu$. This parameter is
obtained by inverting the relation between $r$ and $\mu$ in the non-conserving model, which is computed in the next section.
In order to draw the first order transition line in the conserving model one has to know the large deviations function of the density profile, which has been derived only in the limit
of $a_0\ll1$ \cite{Clincy2003} or $\beta\gg1$ \cite{Bodineau2008}.
However, using the steady-state density profile of the conserving model, discussed in \sref{sec:fol}, one can draw the stability limits around the first order transition line,
 denoted by dashed lines in \fref{fig:MuT_non-equal}a.

\subsubsection{Slow non-conserving dynamics}
A similar critical expansion can be carried out in the non-conserving model. The main difference between the two expansions is the
 additional term in the equation of $b_m$ due to the non-conserving dynamics. %, whose conserving part is given in \eref{eq:complex_bm}.
 In the limit of $p\sim L^{-\gamma}$ and $\gamma>2$ this additional term is negligible with respect to the diffusive term in \eref{eq:complex_bm} for $m\neq 0$.
 This implies that the profile of the vacancies remains flat.
 The non-conserving term is of leading order only for $b_0$, whose equation can be obtained
 by summing \eref{eq:meanfieldB} over $\alpha=A,B,C$ and integrating the result over $x$, yielding
\begin{eqnarray}
\label{eq:b0_non-conserving}
\fl \partial_{\tau} b_0 &= & 3 p L^2 \int_0^1 dx \left ( \rho_0^3 -e^{-3\beta \mu}\rho_A\rho_B\rho_C\right)  \\
\fl &=&  3 p L^2 \Big\{ (1-b_0)^3-e^{-3\beta \mu} \big[ \frac{b_0^3}{27}-\frac{b_0}{3}  \sum_{m_1} a_{m_1}a^\star_{m_1} \nonumber  \\
&&+\frac{1}{3\sqrt{3}} \sum_{m_1,m_2} (a_{m_1}a_{m_2}a_{-m_1-m_2} + c.c.) \big] \Big\}. \nonumber
\end{eqnarray}

In the expansion of $b_0$ around its average value it is useful to consider fluctuations in the overall density,
 defined as $\delta r \equiv b_0- \rs$. In the non-conserving model, $\rs$ denotes the overall density in the homogenous phase,
 obtained by setting $a_m=0$ for $m\neq0$ in \eref{eq:b0_non-conserving}, which yields
\begin{equation}
\label{eq:rmu}
e^{-3\beta\mu}=\frac{27(1-\rs)^{3}}{\rs\left(\rss-9|a_{0}|^{2}\right)}.
\end{equation}
The fluctuations in $b_0$ modify the evolution of $a_m$, which is given by
\begin{eqnarray}
\label{eq:complex_am_full1}
\fl
\partial_{\tau}a_{m}&=&-\frac{2\pi m}{3}\Big[(6\pi m-\sqrt{3}\beta \rs)a_{m} -\sqrt{3} \beta \delta r a_m \nonumber \\
 && +3\beta\sum_{m'=-\infty}^{\infty}a_{-m-m'}^{\star}a_{m'}^{\star}\Big].
\end{eqnarray}
By inserting the definition $b_0=\rs+\delta r$ into \eref{eq:b0_non-conserving} and expanding it in leading order in $\delta r$ one obtains that
slightly below the critical line $\delta r \sim |a_m|^2$. This implies that  $\delta r$ does not affect the linear stability of $a_m$ in \eref{eq:complex_am_full1}, leading to a non-conserving critical line that is identical to the conserving line in \eref{eq:bc_cons_non-equal}. The effect of $\delta r$ appears only in higher order critical points.
 As in the conserving model, equation \eref{eq:complex_am_full1} is expanded in \aref{sec:high_order} in powers of the amplitude of the first excited eigenvector, $\varphi$, yielding
\begin{equation}
\partial_{\tau} \varphi =  4\pi^2 \delta \varphi + G^{nc}_4(\rs,a_0) \varphi \left| \varphi\right|^2 + G^{nc}_6(\rs,a_0) \varphi \left| \varphi\right|^4 + O(\varphi^7).
\end{equation}
In this expansion we find that  $G^{nc}_4(\rs,a_0)\neq G^{c}_4(\rs,a_0)$, leading to a different tricritical points in the
conserving and non-conserving models. % and hence ensemble inequivalence.

In \aref{sec:high_order} we also study numerically $G_6^c(r,a_0)$ and $G_6^{nc}(\rs,a_0)$.
In general, if this coefficient is positive at the point where $\beta=\beta_c$ and $G_4=0$, the model exhibits a tricrtical point connecting the second order and first order transition lines.
 This is the case in the conserving model, plotted
 in \fref{fig:MuT_non-equal}a, and in the non-conserving model only for relatively high values of $|a_0|$. \Fref{fig:MuT_non-equal}b shows the opposite case, where $G_6^{nc}<0$ , which
 occurs in the non-conserving model for $|a_0| \neq 0 $ but relatively small.  In this case the point where $\beta=\beta_c$ and $G_4^{nc}=0$ (TCP in \fref{fig:MuT_non-equal}b)
   is preempted by a critical-end-point (CEP in \fref{fig:MuT_non-equal}b), which connects the second and the first order transition lines.
   The first order transition line continues
   into the ordered phase, where it signifies a transition between a low density and high density ordered phases.
   The first order transition line in this figure is derived in the next section.
    Remarkably, in the equal-densities case the second and first order transition lines are connected by a forth order critical point where $\beta=\beta_c$, $G_4^{nc}=G_6^{nc}=0$ and $G_8^{nc}>0$ \cite{Lederhendler2010b}.
   It is not yet clear whether this high order criticality is merely a coincidence or a result of
   an unidentified symmetry of the model.

\subsection{Nonperturbative study of the non-conseving phase diagram}
\label{sec:fol}

In this section we derive the first order transition line of the ABC model with slow non-conserving dynamics, shown in \fref{fig:MuT_non-equal}b.
The derivation is based on the exact expression for the steady-state density profile of the conserving ABC model, $\boldsymbol \rho^\star(x,\beta,r,a_0)$, which has been computed for $a_0=0,r=1$ in \cite{Ayyer2009}, for $a_0\neq 0,r=1$ in \cite{Cohen2011a} and for $a_0\neq0,0<r\leq1$ in \cite{Cohen2012}.
We first discuss the derivation of $\boldsymbol \rho^\star$ and then use it to compute the large deviations function of $r$.

 In general, the derivation of $\boldsymbol \rho^\star$ is done by setting $\partial_t \rho_\alpha=0$ in \eref{eq:meanfieldA}
 (moving steady-state profiles has been excluded in \cite{Bertini2012}) and integrating over $x$, yielding
 \begin{equation}
 \label{eq:mean_field_int}
 J_\alpha = -\beta \left[\rho_\alpha\left(\rho_{\alpha+1}-\rho_{\alpha+2}\right)\right]-\partial_x \rho_\alpha.
 \end{equation}
Here ${\bf J}\equiv(J_A,J_B,J_C)$ are the steady-state currents of particles, which obey $J_A+J_B+J_C=0$ due to the local particle-conservation.
Through several additional algebraic manipulations, described in \cite{Cohen2011a} for $r=1$,
 the above equation can be reduced to a single ordinary differential equation, which corresponds to the motion of a particle in a quartic potential with $x$ playing the role of time.
 The trajectory of the particle, which corresponds to $ \boldsymbol \rho^\star(x,\beta ,1,a_0)$ can be written in terms of elliptic functions \cite{Cohen2011a}.
Using the mapping discussed in the paragraph below \eref{eq:bc_equal} the profile for arbitrary $r$ can be written as
\begin{equation}
 \label{eq:rho_star}
\boldsymbol \rho^\star(x,\beta,r,a_0)  = r \boldsymbol \rho^\star(x,\beta r,1,a_0/r).
\end{equation}
For brevity, we omit from here on the dependence of $\boldsymbol \rho^\star$ on $\beta$ and $a_0$ and denote the steady-state profile in \eref{eq:rho_star} by $\boldsymbol \rho^\star(x,r)$.

\begin{figure}
\noindent
\begin{centering}\includegraphics[scale=0.58]{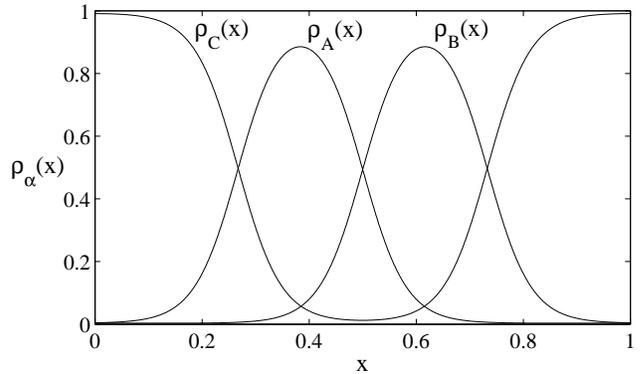}\par\end{centering}
\caption{
\label{fig:profile}
The $m=1$ ordered density profile obtained from the solution of \eref{eq:mean_field_int}
for $r_A=r_B=7/30$ and $r_C=16/30$ and $\beta=25$.}
\end{figure}

According to the analysis presented in \cite{Cohen2011a}, one finds that for low values of $\beta$ the particle has only a constant trajectory which corresponds to a homogenous profile. Whenever $\beta$ is increased above $m \beta_c$ for $m=1,2,\dots $,
with $\beta_c$ defined in \eref{eq:bc_cons_non-equal}, the equation exhibits an additional stable trajectory with $m$ oscillations in the period $0\leq x\leq 1$.
An example of the corresponding profile with $m=1$ is plotted in \fref{fig:profile}. In the equal-density case, the $m=1$ can be shown to exhibit a lower free energy than all the $m>1$ solutions \cite{Ayyer2009}. It is therefore the only ordered profile observed in the thermodynamic limit.
This is assumed to remain true also for arbitrary densities based on numerical evidence from Monte Carlo simulations. %the physical intutition whereby if the $m=2$ is preferable over the $m=1$

We now derive the large deviations function of $r$ using $\boldsymbol \rho^\star(x,r)$ in the limit of very slow non-conserving dynamics, $p\sim L^{-\gamma}$ and $\gamma>2$.
 In this limit the non-conserving term in \eref{eq:meanfieldB} is negligible in comparison to the drift and diffusion terms. Summing \eref{eq:meanfieldB} over $\alpha=A,B,C$ yields
\begin{equation}
\partial_\tau r = 3L^2 p \int_0^1\left(\rho_0^3-e^{-3\beta\mu}\rho_A\rho_B\rho_C\right),
\end{equation}
which implies that $r$ evolves on a time-scale of order $1/p$ (recall that $\tau=t/L^2$).
 The probability to observe a rare value of $r$ can be derived by noting that on a time-scale of $1/p$ the density profile is given by $\boldsymbol \rho^\star(x,r)$.
The master equation of $P(r)$ can therefore be approximated as
\begin{eqnarray}
\label{eq:Master}
  \frac{1}{p}\partial_{t}P(r,t) &=&w^{-}(r+\frac{3}{L}) P(r+\frac{3}{L},t)  + w^{+}(r-\frac{3}{L}) P(r-\frac{3}{L},t) \nonumber \\
& & - P(r,t) \left[ w^{-}(r) + w^{+}(r)\right],
\end{eqnarray}
where the $w^{-}$ and $w^{+}$ denote the average evaporation and deposition rates, respectively.
The evaporation rate is defined using the conserving probability measure, $P(\boldsymbol \zeta \, ; N_0)$, as
\begin{eqnarray}
\label{eq:n_rates1}
w^{-}(r) & \equiv & \frac{1}{L} \, e^{-3\beta\mu} \sum_{\boldsymbol \zeta}  P \left(\boldsymbol \zeta \, ; L-Lr \right) \sum_{i=1}^{L}
A_i B_{i+1} C_{i+2} \\
&=&   \, e^{-3\beta\mu} \int_0^1 \!\! dx\rho_A^\star\left(x,r\right)\rho_B^\star\left(x,r\right)\rho_C^\star\left(x,r\right)+ \mathcal{O}(\frac{1}{L}).\nonumber
\end{eqnarray}
where the first index of $\zeta_{i,\alpha}$ runs cyclically over $i\in [1,L]$. %In the above sum over $i$ we use the projection operators defined in \eref{eq:projection_op}.
 The fact that the leading order contribution to the local correlation function, $A_iB_{i+1}C_{i+2}$, is given by
 the corresponding product of the coarse-grained density profiles is a characteristic property of
diffusive systems. It is shown explicitly for the equal-densities ABC model in \cite{Clincy2003,Ayyer2009} and argued to be true also for arbitrary average densities in \cite{Ayyer2009,Bertini2012}.
The deposition rate is computed in a similar way, yielding
\begin{equation}
\label{eq:n_rates2}
w^{+}(r) = \int_0^1 \!\! dx\left[\rho_0^\star\left(x,r\right)\right]^3+ \mathcal{O}(\frac{1}{L}) = (1-r)^3+ \mathcal{O}(\frac{1}{L}).
\end{equation}
Here we used the fact that the inert vacancies have a flat steady-state profile, $\rho_0^\star\left(x,r\right)=1-r$.

\begin{figure}[t]
\noindent
\begin{centering}\includegraphics[scale=0.58]{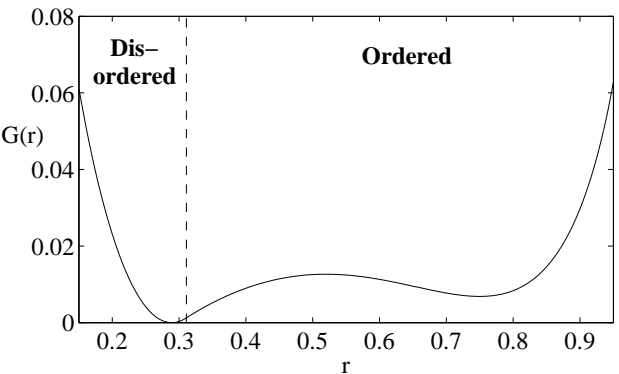}\par\end{centering}
\caption{ \label{fig:Fmu} The large deviations function of the overall density, $r$, for $\beta=40$, $r_A=r_B=r/3-0.025$ and $\mu=-0.0515$.
 For this choice of parameters $\mathcal{G}(r)$ has two local minima, corresponding to a disordered and an ordered phase.
 The dashed line denotes the critical value of $r$, given in \eref{eq:bc_cons_non-equal}.}
\end{figure}

\Eref{eq:Master} corresponds to a one-dimensional random walk in $r$ in the presence of a local potential which scales linearly with $L$.
This implies that its steady-state solution obeys a large deviations principle of the form
\begin{equation}
\label{eq:PN}
P(r) = e^{L  \mathcal{G}\left(r\right)},
\end{equation}
where $ \mathcal{G}$ is the corresponding large deviations function. Inserting this form into \eref{eq:Master} and expanding to first order in $L$ yields
\begin{equation}
\label{eq:Gr_eq}
\left(e^{-3\mathcal{G}'(r)}-1\right)w^{+}(r)+\left(e^{3\mathcal{G}'(r)}-1\right)w^{-}(r)=0.
\end{equation}
This solution to the above equation is given by
\begin{equation}
\label{eq:Gr1}
\mathcal{G}(r)=-\frac{1}{3}\int^r_{r_0} dr' \log \left(w^{-}(r')/ w^{+}(r')\right),
\end{equation}
where $r_0$ is chosen arbitrarily to be the minimal possible particle density, $r_{0}=r-3 \min_{\alpha} \left( r_{\alpha}\right)$.
The large deviations function, $\mathcal{G}(r)$, which is in fact proportional to the potential felt by the random walker, is plotted in \fref{fig:Fmu} for a typical point in parameter-space.

In writing \eqs(\ref{eq:Master})-(\ref{eq:n_rates2}) it appears that the system relaxes to the conserving steady-state profile (on a time-scale $\tau \sim L^2$)  between every microscopic non-conserving dynamical step.
However, since evaporation and deposition occur on $\mathcal{O}(L)$ sites, the average time between each such event scales as $\frac{1}{Lp}\sim L^{\gamma-1}$.
This seems to suggests that \eqs(\ref{eq:Master})-(\ref{eq:n_rates2}) are valid only for $\gamma > 3$.
However, \eref{eq:meanfieldA} implies that on the diffusive time scale, $\tau\sim L^2$, the density profile is given by
\begin{equation}
\boldsymbol \rho(x,t)=\boldsymbol \rho(x,r(t))+\mathcal{O}(pL^2),
\end{equation}
where $r(t)=\sum_\alpha \int_0^1dx \rho_\alpha(x,t)$. This is correct even for $r(t)$ which is far from the steady-state value. Thus as long as $pL^2 \ll 1$, one
would expect the density profile to be given by the steady-state profile of the conserving system with overall density $r(t)$.
This suggests that the dynamics of $r$ described in \eqs(\ref{eq:Master})-(\ref{eq:n_rates2}) is valid whenever $pL^2 \ll 1$, namely for $\gamma>2$.
Note though that for $2<\gamma<3$ the corrections terms in  \eqs(\ref{eq:n_rates1})-(\ref{eq:n_rates2}) scale as $pL^2$ rather than $1/L$.
This argument is supported by our analysis of the $p=\phi L^2$ limit in \sref{sec:fast_fol},
where $\mathcal{G}(r)$ is shown to be given by \eref{eq:Gr1} in for $\phi\to0$.

The large deviations function can be written in a more physically meaningful way as,
\begin{equation}
\label{eq:LDF}
\mathcal{G}(r) =  \beta\mu r -\beta \int_{r_0}^r dr' \mu \left( r' \right),
\end{equation}
where $\mu(r)$ is defined as
\begin{eqnarray}
\label{eq:mu_def}
\mu(r) &= \frac{1}{3\beta} \Big[ & \log\big( \int_0^1dx
\rho_{A}^\star(x,r)\rho_B^\star(x,r)\rho_C^\star(x,r) \big) \nonumber \\
&& - 3 \log\left( 1-r \right) \Big].
\end{eqnarray}
In \cite{Cohen2012}, the function $\mu(r)$ is shown to correspond to the chemical potential of the conserving model,
as measured by a microscopic meter, coupled to it through the dynamics defined in \eref{eq:addremoverates}.
This observation allows one to compare
the phase diagrams of the conserving and non-conserving models on the same axes, as shown in \fref{fig:MuT_non-equal}.
It is important to note that this definition of the chemical potential is dynamics-dependent. In \cite{Cohen2012}, it is
demonstrated how $\mu(r)$ takes a different form depending on the type of non-conserving process, which may not lead to
ensemble inequivalence

\begin{figure}[t]
\noindent
\begin{centering}\includegraphics[scale=0.6]{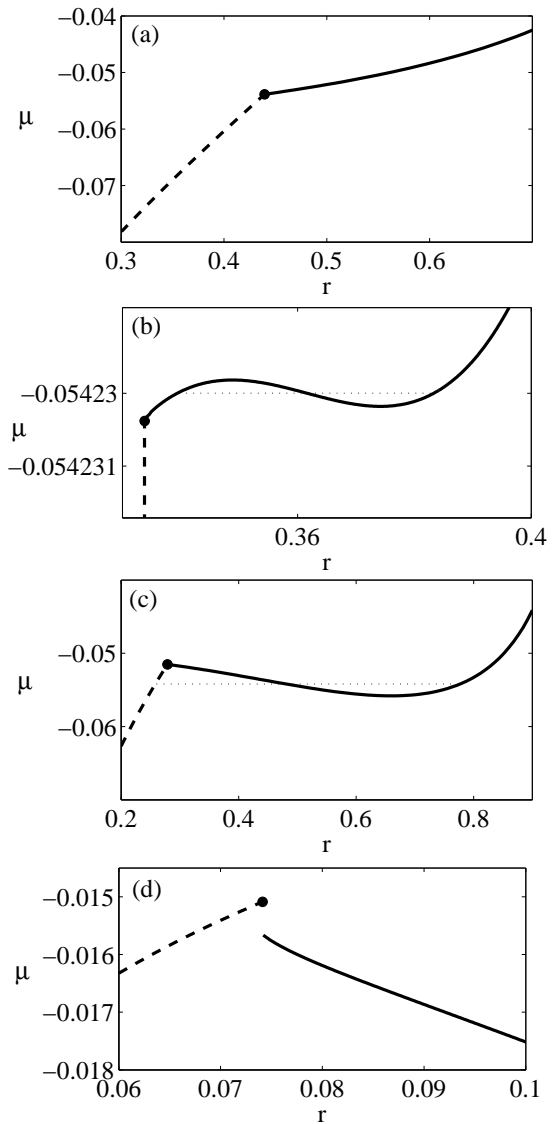}\par\end{centering}
\caption{\label{fig:mu_r} Chemical potential calculated in the conserving model for
the case of $r_A=r_B=r/3-0.01$. Figures (a),(b),(c),(d) correspond to $\beta=25,33.15,40,250$, respectively.
The critical point is denoted by $\bullet$. First order transition points, computed via Maxwell's construction,
 are denoted in (b) and (c) by a dashed lines. The vertical scale in (b) is chosen to be extremely small in order to
 display to the variations in $\mu(r)$ around the first order transition  point between the two ordered phases.  }
\end{figure}

The phase diagram of the conserving and non-conserving models can be studied by analyzing the behaviour of $\mu(r)$ for various values of $\beta$ and $a_0$.
At high temperatures ($1/\beta$), $\mu(r)$ is monotonous but displays a discontinuity in its first derivative, corresponding to
a second order transition point, as shown in \fref{fig:mu_r}a.
 Below the non-conserving tricritical temperature, derived in the previous section, one finds a region of $r$ where $\mu(r)$ can take 3 different values, as shown in \fref{fig:mu_r}c.
The intermediate density solution has negative compressibility and it is thus stable only in the conserving model.
 The non-conserving model exhibits a first order transition
between the homogeneous (right most) and ordered (left most) solutions, which can be located using Maxwell's construction (dashed line).
As discussed in the previous section, in the non-conserving model and in the case where $|a_0|$ is relatively small (presented in \fref{fig:MuT_non-equal}b),
 the second order transition line turns into a first order transition
line at a critical-end-point (CEP). For temperatures slightly higher than the CEP, shown in the inset of \fref{fig:MuT_non-equal}b, there is a second order
transition between the disordered and the order phases as well as a first order transition between two ordered phases.
The nature of this first order transition line can be understood from \fref{fig:mu_r}b, which displays a first order
transition between a low density and a high density ordered phases. The corresponding transition point can also be located using Maxwell's construction (dashed line).
 At very low temperatures, below the conserving tricritical point (\ref{eq:tcl_cons}), $\mu(r)$ exhibits a jump at the critical point, as shown \fref{fig:MuT_non-equal}d.
 This discontinuity is an indication of a first order transition, analyzed in \cite{Cohen2011a}, between a homogenous profile an ordered profile with a finite amplitude.
The values of $\mu$ on the two sides of the discontinuity define two stability limits that are denoted by the dashed lines in \fref{fig:MuT_non-equal}a.

The above derivation of $\mathcal{G}(r)$ is rather general and can be applied to any stochastic model which exhibits a fast conserving dynamics and a slow non-conserving dynamics.
  One may consider
several types of non-conserving dynamical steps of the same variable and or steps of different sizes. In these cases one
can still obtain a general solution for \eref{eq:Master}, independent of the details of $w_-$  and $w_+$, that has a more complicated form
than \eref{eq:Gr1}.
 Another possibility is to consider several non-conserving quantities. In this case the large deviations function of the slow variables
 is given by the steady-state solution of a higher-dimensional random-walk which does not have a general form such as \eref{eq:Gr1}.

\section{ABC model with non-equal densities and comparably slow non-conserving dynamics ($p\sim L^{-2}$)}
\label{sec:fast_evaportation}

In the previous section the non-conserving ABC model was analyzed in the limit $p\sim L^{-\gamma}$ and $\gamma>2$, where its the steady-state profile and the large deviations function of $r$ can be computed analytically. In this section we study how these result are modified in the limit where
\begin{equation}
p=\phi L^{-2},
\end{equation}
and $\phi$ is a parameter that does not scale with $L$.

In \sref{sec:fast_critical} we demonstrate that the conserving and the non-conserving models exhibit for any finite $\phi$ different ordered profiles, as well as different critical lines.  This additional inequivalence can be attributed to the sensitivity of non-equilibrium systems to the details of their dynamics rather than to the existence of long-range correlations.
Remarkably, slightly below the critical line the non-conserving model exhibits a steady-state profile with a non-vanishing drift velocity.

In \sref{sec:fast_fol} we derive an expansion in small $\phi$ of the large deviations function of $r$ using the macroscopic fluctuation theory. The leading order term in the expansion
corresponds to the large deviations function, $\mathcal{G}(r)$, obtained for $\gamma>2$ in \eref{eq:Gr1}. This confirms that $\gamma>2$
 is indeed the limit where the conserving and the non-conserving dynamics are well-separated.
 In the homogenous phase, the large deviations function in \eref{eq:Gr1} is found to be
valid for all values of $\phi$. This conclusion is argued to be correct for a wide class of driven-diffusive models that exhibit a homogenous density profile.

\subsection{Density profile, drift velocity and critical line}
\label{sec:fast_critical}

For $p=\phi L^{-2}$ the dynamics of the density profile in the non-conserving ABC model (\ref{eq:meanfieldB}) can be written as
 \begin{eqnarray}
 \label{eq:meanfieldC}
\partial_{\tau} \rho_\alpha &=& \beta \partial_x\left[\rho_\alpha\left(\rho_{\alpha+1}-\rho_{\alpha+2}\right)\right]+ \partial_x^2\rho_\alpha \\
&&+ \phi \left(\rho_0^3-e^{-3\beta\mu}\rho_A\rho_B\rho_C\right).\nonumber
\end{eqnarray}
Although an analytic steady-state solution for the above equation is not available, it can easily be shown to be different from $\rho^\star(x,r)$, obtained
in the conserving model ($\phi=0$). As mentioned above, this inequivalence between the conserving and the non-conserving models
 is not necessarily due to the existence of long-range correlations in the system.

\Eref{eq:meanfieldC} exhibits a homogenous solution around which it can be expanded.
As demonstrated below, in this case one has to consider a moving density profile and thus use an expansion similar to \eref{eq:rho_expand} with $x$ replaced by $x-vt$, yielding
\begin{eqnarray}
\label{eq:rho_expandC}
\fl \left(\begin{array}{c}
\rho_{A}\\
\rho_{B}\\
\rho_{C}
\end{array}\right)&=&\sum_{m=-\infty}^{\infty}e^{2\pi im(x-v\tau)}\Big[\frac{a_{m}(\tau)}{\sqrt{3}}\left(\begin{array}{c}
1\\
e^{-2\pi i/3}\\
e^{2\pi i/3}
\end{array}\right) \\
&&+\frac{a_{-m}^{\star}(\tau)}{\sqrt{3}}\left(\begin{array}{c}
1\\
e^{2\pi i/3}\\
e^{-2\pi i/3}
\end{array}\right)+\frac{b_{m}(\tau)}{3}\left(\begin{array}{c}
1\\
1\\
1
\end{array}\right)\Big]. \nonumber
\end{eqnarray}
Close to the critical line, we assume that  $a_m,b_m \ll 1$ for all  $m\neq 0$. It is important to note
that the velocity, $v$, does not in general vanish on the critical line. By contrast, in the conserving model the existence of a moving
density profile has been excluded for all values of $\beta$ and $a_0$ in \cite{Bertini2012}.

The main difference between the critical expansions for $\gamma<2$ and for $\gamma=2$ is that in the latter case the $b_m$ amplitudes
 have a drift term in addition to the diffusive term. The drift term, which scales as $\phi$, yields to leading order in $a_m,b_m$ the following equation:
\begin{eqnarray}
\partial_{\tau}b_{m}&=&(2\pi i m v -4\pi^2 m^{2})b_{m} + 3\phi\Big\{ -3(1-b_0)^2 b_{m} \nonumber \\
 && - e^{-3\beta\mu}\big[\frac{b_{0}^{2}b_m}{9} -  \frac{2 b_{0}}{3}(a_{0}a_{-m}^{\star}+a_{m}a_{0}^{\star})  \\
 &&+\frac{1}{\sqrt{3}}(a_{0}^{2}a_{m}+a_{0}^{\star2}a_{-m}^{\star})\big]\Big\} +\mathcal{O}(a_{m}^{2}).\nonumber
 \end{eqnarray}
Assuming a steady state, $\partial_{\tau}b_{m}=0$, and considering $\beta$ that are slightly below the critical line, where $b_m,a_m \ll 1$, one obtains
 a linear dependence between $b_m$ and $a_m$, given by
 \begin{equation}
\label{eq:bm_kappa}
 b_{m}=\kappa_m  a_{m}+\kappa^{\star}_m a_{-m}^{\star}+\mathcal{O}(a_{m}^{2}),
 \end{equation}
where
 \begin{equation}
 \label{eq:kappa}
\kappa_m \equiv  \phi \frac{3e^{-3\beta \mu}(\sqrt{3}a_{0}^{2}-2 ra_{0}^{\star})}{12\pi^2 m^{2}-6\pi i m v+27 \phi(1-r)^2+ \phi r^{2}e^{-3\beta\mu}}.
 \end{equation}
As expected, $\kappa_m $ vanishes in the $\phi\to0$ limit as well as for equal-densities ($a_0=0$).

In order to complete the expansion one has to consider the evolution of $a_m$, obtained by the expansion of \eref{eq:meanfieldC} in small $a_m$ and $b_m$.
The non-conserving term in  \eref{eq:meanfieldC} does not enter directly into the equation of $a_m$ since it is proportional to the vector $(1,1,1)^T$, whereas $a_m$ is the amplitude of
the orthogonal vector $(1,e^{-2\pi i /3},e^{2\pi i /3})^T$. The leading order effect of the non-conserving dynamics comes from the term that couples $b_m$ and $a_0$. Expanding
\eref{eq:meanfieldC} using the form in \eref{eq:rho_expandC} and the expression for $b_m$ in \eref{eq:bm_kappa} yields to leading order
\begin{equation}
\label{eq:dam2}
\left(\begin{array}{c}
\partial_{\tau}a_{m}\\
\partial_{\tau}a_{-m}^{\star}
\end{array}\right)=-\frac{2\pi m}{3}A_{m}\left(\begin{array}{c}
a_{m}\\
a_{-m}^{\star}
\end{array}\right)+\mathcal{O}(a_{m}^{2}),
\end{equation}
where $A_{m}$ is given by
\begin{eqnarray}
 A_{m}&=&\left(\begin{array}{cc}
6\pi m-\sqrt{3}\beta r-3iv-\sqrt{3}\beta a_{0}\kappa_m  &  \\
-6\beta a_{0}+\sqrt{3}\beta a_{0}\kappa_m  &
\end{array}\right. \\
&&\left. \begin{array}{cc}
 \qquad & 6\beta a_{0}^{\star}-\sqrt{3}\beta a_{0}^{\star}\kappa^{\star}_m \\
 \qquad & 6\pi m+\sqrt{3}\beta r-3iv+\sqrt{3}\beta a_{0}^{\star}\kappa^{\star}_m
\end{array}\right). \nonumber
\end{eqnarray}
It is easy to see that for $k_m=0$ and $v=0$ the $A_{m}$ matrices reduce to those in \eref{eq:complex_am}.

\begin{figure}[t]
\noindent
\begin{centering}\includegraphics[scale=0.58]{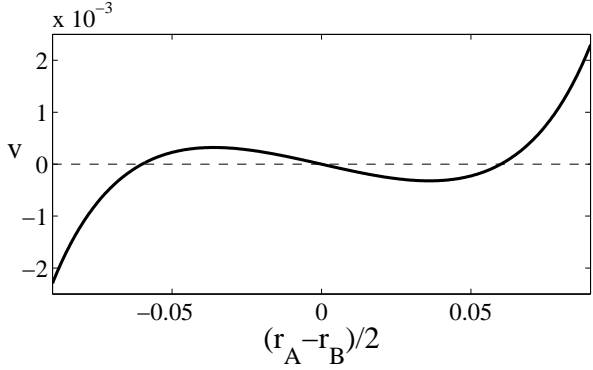}\par\end{centering}
\caption{\label{fig:speed} Velocity of the unstable mode on the critical point for $r_A=0.18+x$, $r_B=0.18-x$, $r_C=0.24$ and $\phi=1$. The horizontal
axis represents the value of $x$.}
\end{figure}

The critical point is defined by the lowest temperature ($1/\beta$) for which $\det A_m=0$ for one of the $m$'s.
The explicit form of this condition is given by
\begin{eqnarray}
\label{eq:detAm}
&9(2 m \pi-iv)^2- 6\sqrt{3} \beta (2 i m\pi+v) \Im[a_{0}\kappa_m] \\\nonumber
&-3\beta^{2}\left(r^{2}-12|a_{0}|^{2}+2r \Re[a_{0}\kappa_m ]+4\sqrt{3}|a_{0}|^{2}\Re[\kappa_m ]\right)=0,
\end{eqnarray}
where $\Re$ and $\Im$ denote the real and imaginary parts, respectively.
In general  $\Im[a_{0}\kappa_m ]\neq 0$ and therefore the solution of \eref{eq:detAm} yields $v\neq0$.
The velocity, $v$, is given by the solution of a cubic equation and is thus omitted for the sake of brevity. This solution is studied
 below in the limit of $\phi \ll 1$ and plotted for a specific line in parameter space in \fref{fig:speed}.

In the special case where $a_0 = |a_0|e^{\pi i \ell /3}$ for $\ell =0,1,2$, one finds from \eref{eq:kappa} that $\Im[a_{0}\kappa_m ]\propto v$,
 which in turn leads to a solution of \eref{eq:detAm} with $v=0$.
This condition can be shown to correspond to the case where
at least two of the species have the same densities. This result conforms with the intuition that in the two-equal-densities case the system does not have a preferred drift direction and hence
$v=0$.  The vanishing of the velocity is evident in \fref{fig:speed} at $r_A=r_B$, $r_A=r_C$ or $r_B=r_C$.
In the two-equal-densities case the fact that $\Im[a_{0}\kappa_m ]=0$ simplifies \eref{eq:detAm}.
As in the previous sections the critical $\beta$ is obtained when the $m=1$ Fourier mode becomes unstable. The solution of \eref{eq:detAm} with
$m=1$ and $\Im[a_{0}\kappa_m ]=0$ yields the following critical line in the two-equal-densities case:
 \begin{equation}
\beta_c=\frac{2\pi \sqrt{3}}{\sqrt{r^{2}-12|a_{0}|^{2}+2r a_{0}\kappa_1 +4\sqrt{3}|a_{0}|^{2}\Re[\kappa_1 ]}}.
\end{equation}
As expected, in the $\phi\to0$ limit, where $\kappa_1 \to0$, this line coincides with the critical line in \eref{eq:bc_cons_non-equal} obtained for $\gamma>2$.

For arbitrary densities, the drift velocity can be studied analytically in the $\phi\ll1$ limit. In this case, one would expect $v$ to vanish with $\phi$, and therefore
\eref{eq:bm_kappa} can be written as
 \begin{equation}
 \label{eq:kappa1}
\kappa_m = \phi\frac{\sqrt{3}a_{0}^{2}-2 ra_{0}^{\star}}{4\pi^2 m^2}e^{-3\beta\mu}+o(\phi).
 \end{equation}
Similarly to the $\gamma>2$ case, the critical transition occurs for $\phi\ll 1$ when the $m=1$ Fourier mode becomes unstable.
Inserting \eref{eq:kappa1} into \eref{eq:detAm} for $m=1$ yields
 \begin{equation}
  \label{eq:v_small_phi}
v = - \frac { \phi }{ \sqrt{3}}\Im[a_{0}\kappa_1 ] \beta_c+\mathcal{O}(\phi^2), % =- \frac { \phi }{ 4\pi^2}\Im[ a_{0}^3] \beta_c e^{-3\beta\mu}+\mathcal{O}(\phi^2),
 \end{equation}
 where $\beta_c$ denotes the critical $\beta$ of the conserving model, given in \eref{eq:bc_cons_non-equal}.
From \eqs(\ref{eq:kappa1}) and (\ref{eq:v_small_phi}) one obtains that $v \propto \phi \Im[ a_{0}^3]$, which implies
 the drift velocity vanishes cubically deviation from the equal-density point,  $|a_0|$.
\Eref{eq:v_small_phi} can also be expressed in terms of $r_A,r_B$ and $r_C$ using the fact that $\Im[ a_{0}^3]=-\frac{1}{2}(r_A-r_B)(r_B-r_C)(r_C-r_A)$ and $e^{-3\beta \mu}=(1-r)^3/r_Ar_Br_C$, which
yields
 \begin{equation}
v = \phi \beta_c \frac{(1-r)^3(r_A-r_B)(r_B-r_C)(r_C-r_A)}{8\pi^2 r_A r_B r_C} +\mathcal{O}(\phi^2).
 \end{equation}
 The latter expression indeed vanishes in the two-equal-densities case.

\subsection{Corrections to the large deviations function of $r$}
\label{sec:fast_fol}
In this section we study the large deviations properties of the non-conserving ABC model in the limit where $p = \phi L^{-2} $. In this limit the probability of a rare
density profile can be expressed using the macroscopic fluctuation theory by extremizing the action of the model over all the possible instanton paths leading to the target profile
\cite{freuidlin1994random,bertini2001fluctuations,jordan2004transport,bertini2005current,tailleur2008mapping,touchette2012large}.
Here we specifically consider events where a rare number of particles is observed. We show that in the $\phi \to 0$ limit these events are realized by an instanton that passes only through steady-state profile of the conserving
system, i.e. ${ \boldsymbol \rho}(x,\tau) = { \boldsymbol \rho}^\star(x,r(\tau))$ where $r(\tau)$ can be computed analytically. The large deviations function of $r$, obtained using this instanton path, coincides in the $\phi \to 0$ limit
 with the one computed for $p\sim L^{-\gamma}$ with $\gamma>2$ in \sref{sec:fol}. In the homogenous phase we find that this instanton path and the corresponding large deviations function remain valid for all $\phi$.
 In the ordered phase we write an expansion of the instanton in small $\phi$ and derive equations that describe its first order correction.
 This expansion is presented in a relatively general form, which can easily adapted to other driven diffusive models.

The macroscopic fluctuation theory deals with the probability to observe a trajectory of
 the macroscopic current of particles within a given time interval, denoted here by $[0,{T} ]$. In our case we consider three conserving currents, ${\bf j}(x,\tau)\equiv (j_A(x,\tau),j_B(x,\tau),j_C(x,\tau))^T$, which result from local exchanges of particles, and a non-conserving current,
  $k(x,\tau)$, which results from evaporation and deposition of triplets of particles. The density profile is obtained from those two currents via the following continuity equation:
\begin{equation}
\label{eq:contin}
\partial_\tau \boldsymbol{\rho}(x,\tau) = -\partial_x \mathbf{j}(x,\tau) + \phi {\bf u} k(x,\tau),
\end{equation}
where ${\bf u} = (1,1,1)^T$. It is important to note that $k(x,\tau)$ is the non-conserving current of triplets of particles and not of single particles.

 The probability of a trajectory in current-space, defined by ${\bf j}(x,\tau)$ and $k(x,\tau)$, is given by the following large deviations principle:
\begin{eqnarray}
\label{eq:Prhojk}
 && Pr[\mathbf{j}(x,\tau),k(x,\tau)|\boldsymbol{\rho}(x,0)] \sim \\
&& e^{-L\int_{0}^{{T}}d\tau\int_{0}^{1}dx\left[\mathcal{L}_{c}(\boldsymbol{\rho}(x,\tau),\mathbf{j}(x,\tau))+\phi\mathcal{L}_{nc}(\boldsymbol{\rho}(x,\tau),k(x,\tau))\right]}, \nonumber
\end{eqnarray}
where $\mathcal{L}_c$ and $\mathcal{L}_{nc}$ denote the conserving and the non-conserving Lagrangian densities, respectively.
These functionals, whose explicit form is given below, correspond to the `probability cost' of the conserving and of the non-conserving noises, respectively.
  The steady-state probability of a density profile $\bar{\boldsymbol \rho}(x)$ can be obtained from \eref{eq:Prhojk}
 by integrating it over all the trajectories for which $\boldsymbol \rho(x,{T})=\bar{\boldsymbol \rho}(x)$ in the limit of ${T}\to \infty$, yielding
\begin{eqnarray}
\label{eq:Prhojk1}
    && Pr[\boldsymbol \rho(x)=\bar{\boldsymbol \rho}(x)]  = \\
  && \lim_{{T}\to \infty}\int D \boldsymbol \rho \bigg|_{\boldsymbol \rho(x,{T})=\bar{\boldsymbol \rho}(x)}  D{\bf j }D k Pr[\mathbf{j}(x,\tau),k(x,\tau)|\boldsymbol{\rho}(x,0)] \nonumber \\
    && \qquad \times \prod_\alpha \delta[ \partial_\tau {\rho}_\alpha(x,\tau) + \partial_x j_\alpha (x,\tau) - \phi k(x,\tau)], \nonumber
\end{eqnarray}
where $\boldsymbol{\rho}(x,0)$ can be chosen arbitrarily.

The conserving Lagrangian density, $\mathcal{L}_c$,  has been derived for the standard ABC model ($p=0,r=1$) in \cite{Bodineau2008}.
It is important to note that this expression cannot be mapped onto the $r<1$ case, as done in \eref{eq:rho_star}
for the steady-state density profile. This is because for $r<1$ one has to  consider fluctuations in $\rho_A(x)+\rho_B(x)+\rho_C(x)$ which were not allowed for $r=1$.
We now sketch the derivation of $\mathcal{L}_c$ for $r< 1$, by following the lines of derivation presented in \cite{Bodineau2008}.
First we consider the Langevin equation for the macroscopic current, ${\bf j}(x,t)$, given by
\begin{equation}
j_{\alpha}(x,\tau)=-\chi_\alpha(\boldsymbol \rho(x,\tau))-\partial_{x}\rho(x,\tau)+\frac{1}{\sqrt{L}}\eta_{\alpha},
\end{equation}
where $\chi_\alpha (\boldsymbol \rho)=\beta\rho_{\alpha}(x,t)[\rho_{\alpha+1}(x,t)+\rho_{\alpha+2}(x,t)]$ denotes the drift term.
The variables $\eta_\alpha$ denote gaussian white noise, whose correlations are given by
\begin{equation}
\left\langle \eta_{\alpha}(x,\tau)\eta_{\alpha'}(x',\tau')\right \rangle= \delta(x-x')\delta(\tau-\tau') \Sigma_{\alpha,\alpha'}(\boldsymbol \rho(x,\tau)),
\end{equation}
where  $\Sigma_{\alpha,\alpha'}(\boldsymbol \rho)$ is the conductivity matrix. The latter can be obtained by calculating the local covariance of the
conserving currents for $\beta=0$
 (see discussion in Appendix B of \cite{Bertini2012} and in Section 2.3 of \cite{Bodineau2008}), yielding
\begin{equation}
\Sigma_{\alpha\alpha'}(\boldsymbol \rho)=\left\{ \begin{array}{ccc}
2\rho_{\alpha}(1-\rho_{\alpha}) & \qquad & \alpha=\alpha'\\
-2\rho_{\alpha}\rho_{\alpha'} & \qquad & \alpha\neq\alpha'
\end{array}\right. .
\end{equation}
The probability of a current trajectory ${\bf j}(x,\tau)$ is obtained by integrating over the probability distribution of the noise variable and performing the Martin-Siggia-Rose procedure.
This yields a large deviations function given by the integral over the following Lagrangian density:
\begin{equation}
\mathcal{L}_{c}(\boldsymbol{\rho},\mathbf{j})=\left[\mathbf{j}+\partial_{x}\boldsymbol{\rho}
+\boldsymbol \chi(\boldsymbol{\rho})\right]^T\Sigma^{-1}(\boldsymbol{\rho})\left[\mathbf{j}+\partial_{x}\boldsymbol{\rho}+{\boldsymbol \chi}(\boldsymbol{\rho})\right].
\end{equation}

 The non-conserving Lagranian density, $\mathcal{L}_{nc}$, was derived for a
 general driven diffusive system in \cite{jona1993large,bodineau2010current,bodineau2012large}. We now repeat the heuristic derivation of $\mathcal{L}_{nc}$, presented for a general
 driven-diffusive model in \cite{bodineau2010current}, in the context of the non-conserving ABC model.
 The local non-conserving current is given by the combination of two Poisson processes: condensation of triplets of particles with rate,
  $\phi \mathcal{C}\big(\boldsymbol{\rho}(x,\tau)\big)$, and annihilation of triplets of particles, with rate $\phi \mathcal{A}\big(\boldsymbol{\rho}(x,\tau)\big)$, where
\begin{equation}
 \mathcal{A}(\boldsymbol{\rho})=  e^{-3\beta\mu}\rho_{A}\rho_{B}\rho_{C},\quad \mathcal{C}(\boldsymbol{\rho})=(1-\rho_{A}-\rho_{B}-\rho_{C})^{3}.
\end{equation}
For a small segment of the system, $dx$, and a small time interval, $d\tau$,
 the probability to observe $k \phi dx d\tau$ triplets of particles added or removed (depending on the sign of $k$) due to the non-conserving processes is given by
\begin{eqnarray}
 && Pr[\int_{x}^{x+dx}\int_{\tau}^{\tau+\phi d\tau}dxd\tau k(x,\tau)=k \phi dxd\tau]= \\
  && \sum_{n=\max(0, L k \phi d\tau dx)}^{\infty}P_{\phi \mathcal{C}(\boldsymbol{\rho})dxdt}^{(\mathrm{Pois.})}(n)P_{\phi \mathcal{A}(\boldsymbol{\rho})dxdt}^{(\mathrm{Pois.})}(n-L k \phi dxdt)\nonumber \\
 && \qquad \sim  e^{-\phi\mathcal{L}_{nc}(\boldsymbol{\rho},k)d\tau dx}, \nonumber
\end{eqnarray}
where  $P_\lambda^{(\mathrm{Pois.})}$ denotes the Poisson distribution function with a mean of $\lambda$.
  The non-conserving action is obtained by performing the saddle point approximation over the
above sum, yielding
\begin{eqnarray}
\fl  \mathcal{L}_{nc}(\boldsymbol{\rho},k)&=&\mathcal{C}(\boldsymbol{\rho})+\mathcal{A}(\boldsymbol{\rho})
-\sqrt{k^{2}+4\mathcal{A}(\boldsymbol{\rho})\mathcal{C}(\boldsymbol{\rho})} \\
&& +k\ln(\frac{\sqrt{k^{2}+4\mathcal{A}(\boldsymbol{\rho})\mathcal{C}(\boldsymbol{\rho})}+k}{2\mathcal{C}(\boldsymbol{\rho})}). \nonumber
\end{eqnarray}
This functional form is identical to the one obtained in  \cite{bodineau2010current} for a general $\mathcal{A}$ and $\mathcal{C}$.
Since the conserving and the non-conserving processes occur independently, the corresponding Lagrangian densities can be added together, yielding \eref{eq:Prhojk}.

The integral in \eref{eq:Prhojk1} can be evaluated using the saddle point approximation, whereby the probability of a density profile is governed by
the action over the most probable path to reach $\bar{\boldsymbol \rho}(x)$ at time ${T}\to \infty$.
In order to find this path it is useful to consider two functions, ${\bf H}(x,\tau)$ and $G(x,\tau)$, defined as
\begin{eqnarray}
\label{eq:H_def}
\Sigma (\boldsymbol \rho) \partial_x {\bf H} = {\bf j} + \partial_x{\boldsymbol \rho} + {\boldsymbol \chi}(\boldsymbol \rho),\\
\label{eq:G_def}
k=\mathcal{C}(\boldsymbol \rho)e^{G} - \mathcal{A}(\boldsymbol \rho)e^{-G}.
\end{eqnarray}
Physically, ${\bf H}(x,\tau)$ and $G(x,\tau)$ correspond to the fields exerted by the external baths that drive the conserving and the non-conserving currents, respectively.

In terms of  ${\bf H}(x,\tau)$ and $G(x,\tau)$ the Lagrangian densities can be written as
\begin{eqnarray}
\label{eq:Hamiltonian_action}
 \mathcal{L}_c(\boldsymbol \rho, H)&=&\frac{1}{2}(\partial_x {\bf H})^T\Sigma(\boldsymbol\rho) \, \partial_x {\bf H}, \\
 \mathcal{L}_{nc}(\boldsymbol \rho, G) &=& \mathcal{C}(\boldsymbol \rho)(1-e^{G}+G e^{G})  \\
 && + \mathcal{A}(\boldsymbol \rho)(1-e^{-G}-G e^{-G}). \nonumber
\end{eqnarray}
By expanding the action,
\begin{equation}
 \label{eq:action}
\mathcal{I}(\boldsymbol \rho, {\bf H},G)\equiv \int_0^{{T}} d\tau \int_0^1dx [\mathcal{L}_c(\boldsymbol \rho, {\bf H}) +  \mathcal{L}_{nc}(\boldsymbol \rho, G)],
\end{equation}
 in small variations in ${\bf H}(x,\tau)$ and $G(x,\tau)$
around the  extremizing trajectory of $\mathcal{I}$, it is shown in \aref{sec:hamilton} that this trajectory obeys the following equations:
\begin{eqnarray}
 \label{eq:Hamil_G}
  G&=&  H_A+H_B+H_C,\\
 \label{eq:Hamil_H}
 \partial_\tau H_\alpha & = &  - \partial_x^2 H_\alpha  + \partial_x {\bf H}^T \partial_{\rho_\alpha} \boldsymbol \chi (\boldsymbol \rho)
  -\frac{1}{2}(\partial_x {\bf H})^T\partial_{\rho_\alpha}\Sigma (\boldsymbol \rho) \, \partial_x {\bf H} \nonumber \\
& &+  \phi [\partial_{\rho_\alpha} \mathcal{C}(\boldsymbol \rho)(1-e^{G}) + \partial_{\rho_\alpha} \mathcal{A}(\boldsymbol \rho)(1-e^{-G})],
\end{eqnarray}
where $\boldsymbol \rho$ can be written in terms of ${\bf H}$ and $G$ using the continuity equation (\ref{eq:contin}), which yields
\begin{eqnarray}
 \label{eq:Hamil_rho}
\partial_\tau \rho_\alpha & =&  \partial_x^2 \rho_\alpha + \partial_x \chi_\alpha(\boldsymbol \rho)-\partial_x[\Sigma(\boldsymbol \rho) \partial_x {\bf H}]_\alpha \\
&& +\phi [\mathcal{C}(\boldsymbol \rho)e^{G} - \mathcal{A}(\boldsymbol \rho)e^{-G}]. \nonumber
\end{eqnarray}
It is useful to note that the derivation of \eqs(\ref{eq:H_def})-(\ref{eq:Hamil_rho}) can be regarded as a transformation into a Hamiltonian picture.
 In this respect, the fields ${\bf H}(x,\tau)$ and $G(x,\tau)$ are the
momenta conjugate to $-\int_0^\tau d\tau' \partial_x {\bf j} (x,\tau')$ and $\int_0^\tau d\tau'  k (x,\tau')$, respectively,
and \eqs(\ref{eq:Hamil_G})-(\ref{eq:Hamil_rho}) play the role of Hamilton's equations.

As discussed above, the solution to \eqs(\ref{eq:Hamil_G})-(\ref{eq:Hamil_rho}) in the $\phi \ll 1$  limit is expected to yield a density profile of the form
${ \boldsymbol \rho}(x,t) = { \boldsymbol \rho}^\star(x,r(\tau))+\mathcal{O}(\phi)$.
Inserting this ansatz into \eref{eq:Hamil_rho} yields ${\bf H}(x,\tau)$ which is constant in space up to deviations proportional to $\phi$. In addition it is
evident that $r(\tau)$ evolves on a slow time-scale defined by $\phi\tau$. One can therefore consider the following expansion of the instanton path:
\begin{eqnarray}
\label{eq:instanton_guess_H}
H_\alpha(x,\tau) &= & \frac{1}{3}g(\phi \tau) + \sum_{k=1}^\infty \phi^k \delta^{(k)}_{H,\alpha}(x,\phi \tau), \\
\label{eq:instanton_guess_rho}
\rho_\alpha(x,\tau)& = & \rho^\star_\alpha(x,r(\phi\tau)) + \sum_{k=1}^\infty \phi^k \delta^{(k)}_{\rho,\alpha}(x,\phi \tau).
\end{eqnarray}

The functions appearing in the above expansion can be computed by inserting \eqs(\ref{eq:instanton_guess_H})-(\ref{eq:instanton_guess_rho})
into \eqs(\ref{eq:Hamil_G})-(\ref{eq:Hamil_rho}) and solving them order by order in $\phi$.
 The leading order terms, $g(s)$ and $r(s)$ (with $s$ denoting $\phi \tau$), are obtained from integrating over $x$ and summing over $\alpha$ the $\mathcal{O}(\phi)$
terms in \eqs(\ref{eq:Hamil_G})-(\ref{eq:Hamil_rho}). This integral and sum of \eref{eq:Hamil_rho} yields
 \begin{eqnarray}
 \label{eq:drs}
 \frac{dr}{d s} &= & 3 \int_0^1dx [ \mathcal{C}(\boldsymbol \rho^\star_s)e^{g(s)} - \mathcal{A}(\boldsymbol \rho^\star_s)e^{-g(s)}],
\end{eqnarray}
where $\boldsymbol \rho^\star_s \equiv \boldsymbol \rho^\star\big(x,r(s)\big)$.
Similarly, multiplying \eref{eq:Hamil_H} by $\partial_r \rho^\star_\alpha(x,r(s))$ and integrating over $x$ yields
 \begin{eqnarray}
 \label{eq:dhs}
\fl  \frac{d g}{d s}& = & \int_0^1dx \sum_\alpha \partial_r \rho^\star_\alpha(x,r(s))  [\partial_{\rho_\alpha} \mathcal{C}(\boldsymbol \rho^\star_s)(1-e^{g(s)}) \nonumber \\
&& + \partial_{\rho_\alpha} \mathcal{A}(\boldsymbol \rho^\star_s)(1-e^{-g(s)})].
\end{eqnarray}
Here all the terms that contain $\delta_{H,\alpha}^{(1)}$ were removed using integration by parts and the fact that $\partial_x^2 \rho_\alpha^\star + \partial_x \chi_\alpha(\boldsymbol \rho^\star)=0$. \Eref{eq:dhs} has a solution of the form,
\begin{equation}
\label{eq:hs_soltion}
g(s) = \ln \Big[\int_0^1dx \mathcal{A}(\boldsymbol \rho^\star_s)\Big] - \ln \Big[\int_0^1dx \mathcal{C}(\boldsymbol \rho^\star_s)\Big],
\end{equation}
which when inserted into \eref{eq:drs} yields
 \begin{eqnarray}
\label{eq:hr_soltion}
\frac{dr}{d s} &= & 3\int_0^1dx \big[ \mathcal{A}(\boldsymbol \rho^\star_s) - \mathcal{C}(\boldsymbol \rho^\star_s)\big].
\end{eqnarray}
In general, since $\boldsymbol \rho^\star(x,r)$ has a rather complicated form, the solution of \eref{eq:hr_soltion} can only be obtained numerically.
Nevertheless, in the following paragraphs we draw some interesting conclusions from the general form of \eref{eq:hr_soltion}.

To leading order in $\phi$, \eref{eq:hr_soltion} describes the time reversal of the relaxation trajectory of $r$ from an atypical initial value.
 The relaxation trajectory can be obtained by inserting a solution of the form of \eref{eq:instanton_guess_rho} into
 \eref{eq:Hamil_rho} with ${\bf H}(x,\tau)=0$ and keeping the leading order in $\phi$.
  This time-reversal symmetry conforms with our intuition that in the limit of very slow non-conserving dynamics, the dynamics
 of $r$ can be described as an equilibrium one-dimensional random walk. Since for $\gamma>2$ we know that this instanton path is the most probable one that realizes a rare value of $r$,
 it is plausible to assume the expansion around it for $\gamma=2$ and $\phi \ll 1$ is also the global minimum of $\mathcal{I}$.

Under this assumption, the probability to observe a rare value of the overall density is obtained by evaluating the action (\ref{eq:action}) over the instanton path in \eqs(\ref{eq:instanton_guess_H})-(\ref{eq:instanton_guess_rho}).
This statement can be written formally as
 \begin{eqnarray}
\fl  P(\bar{r}) &=& P[\boldsymbol \rho(x)=\boldsymbol \rho^\star(x,\bar{r})+O(\phi)]  \\
&  \sim& \lim_{\phi {T}\to\infty} e^{-L \int_0^{\phi {T}} ds \int_0^1 dx  [\mathcal{L}_c(\boldsymbol \rho^\star_s, \frac{1}{3}g(s) ) +  \mathcal{L}_{nc}(\boldsymbol \rho^\star_s,  g(s))] + \mathcal{O}(\phi) }, \nonumber
 \end{eqnarray}
where $g(0)$ is chosen such that $r(\phi {T})=\bar{r}$.
Using \eqs(\ref{eq:hs_soltion})-(\ref{eq:hr_soltion}) the above exponential can be written as
 \begin{equation}
 \label{eq:ldf_AC}
 P(\bar{r})  \sim e^{\frac{L}{3} \int^{\bar r} dr'  \{ \log [ \int_0^1 dx \mathcal{C}(\boldsymbol \rho^\star(x,r')) ] - \log [\int_0^1 dx \mathcal{A}(\boldsymbol \rho^\star(x,r'))] \} + \mathcal{O}(\phi)}.
 \end{equation}
As expected, the leading order term of this large deviations function is  identical to $\mathcal{G}(r)$ obtained for $\gamma>2$ in \sref{sec:fol}.

Corrections to $\mathcal{G}(r)$ can be obtained by calculating the $\mathcal{O}(\phi)$ corrections to the instanton path. Considering the $\mathcal{O}(\phi)$ terms when expanding
 \eqs(\ref{eq:Hamil_H})-(\ref{eq:Hamil_rho}), we find that
\begin{eqnarray}
\label{eq:delta_rho}
  0&=&\partial_{x}^{2}\delta_{\rho,\alpha}^{(1)}+\partial_{x}[\delta_{\rho,\alpha}^{(1)}\partial_{\rho_{\alpha}}\chi_{\alpha}(\boldsymbol{\rho}^{\star}_s)]
-\partial_{x}[\Sigma(\boldsymbol{\rho}^{\star})\partial_{x}\boldsymbol{\delta}_{H}^{(1)}]_{\alpha}\\
&+&\mathcal{C}(\boldsymbol{\rho}^{\star}_s)e^{g(s)}-\mathcal{A}(\boldsymbol{\rho}^{\star}_s)e^{-g(s)}\nonumber \\
  &-& \partial_r  \boldsymbol \rho^\star_s \int_0^1dx  [ \mathcal{C}(\boldsymbol \rho^\star_s)e^{g(s)} - \mathcal{A}(\boldsymbol \rho^\star_s)e^{-g(s)}],  \nonumber \\
\label{eq:delta_H}
0&=&-\partial_{x}^{2}\delta_{H,\alpha}^{(1)}+\partial_{x}\boldsymbol\delta_{H}^{(1),T} \partial_{\rho_{\alpha}}\boldsymbol \chi(\boldsymbol{\rho}^{\star}_s) \\
&+&\partial_{\rho_{\alpha}}\mathcal{C}(\boldsymbol{\rho}^{\star}_s)(1-e^{g(s)})+\partial_{\rho_{\alpha}}\mathcal{A}(\boldsymbol{\rho}^{\star}_s)(1-e^{-g(s)}) \nonumber \\
  &-&  \int_0^1dx \partial_r \boldsymbol \rho^\star_s [ \partial_{\rho_{\alpha}}\mathcal{C}(\boldsymbol{\rho}^{\star}_s)(1-e^{g(s)})+\partial_{\rho_{\alpha}}\mathcal{A}(\boldsymbol{\rho}^{\star}_s)(1-e^{-g(s)})],  \nonumber
\end{eqnarray}
where as before $s\equiv \phi \tau$.
 Since $r(s)$ does not have a simple analytic expression in the ABC model, the above equations can only be solved numerically.

In the homogenous phase, where $\rho^\star_\alpha(x,r) =r_\alpha$, the source terms in \eqs(\ref{eq:delta_rho})-(\ref{eq:delta_H}) vanish, yielding a set of homogenous partial differential equations that exhibit a trivial solution, $\delta^{(1)}_{H,\alpha}=
\delta^{(1)}_{\rho,\alpha}=0$. In fact, in this case the leading order term of the instanton, $H_\alpha(x,\tau)=\frac{1}{3}\,g(\phi\tau)$ and
 $\boldsymbol \rho(x,\tau)=\boldsymbol \rho^\star(x,r(\phi\tau))$, can be shown to satisfy \eqs(\ref{eq:Hamil_G})-(\ref{eq:Hamil_rho}) to all orders in $\phi$.
 This conclusion does not depend on the specific form of $\Sigma(\boldsymbol \rho)$ and
 $\boldsymbol \chi(\boldsymbol \rho)$, and can be shown to remain valid also when considering a density-dependent diffusion coefficient.
 This implies that for any homogenous driven-diffusive system with a
slow particle-non-conserving dynamics ($p \sim L^{-\gamma}$ with $\gamma\geq 2$) that is characterized by space-independent functions $\mathcal{A}(r)$ and $\mathcal{C}(r)$,
the large deviations principle of the overall particle-density is given by
 \begin{equation}
 P(\bar{r})  \sim e^{ \frac{L}{\omega} \int^ {\bar r} dr' [\log \mathcal{C}(r') -\log \mathcal{A}(r')]}.
 \end{equation}
Here $\omega$ are the number of particles added and removed in each condensation and annihilation process. In the non-conserving ABC model $\omega=3$.

Another interesting result of the above expansion concerns the time-reversal symmetry found in the dynamics of $r$, whereby its excitation trajectory to a rare density, $\bar{r}$, is equal to the time-reversal
of the relaxation path which starts from $r=\bar{r}$. This symmetry is broken for the local non-conserving current of particles, $k(x,\tau)$, for systems where $\mathcal{C}(\boldsymbol \rho^\star)$ and
 $\mathcal{A}(\boldsymbol \rho^\star)$ are not homogenous in space. On the one hand, in such cases the relaxation path of $k$, obtained by setting $G=0$ in \eref{eq:G_def}, is given by
  \begin{equation}
 k_{\downarrow}(x,\tau)=\mathcal{C}(\boldsymbol \rho^\star_{\phi \tau})-\mathcal{A}(\boldsymbol \rho^\star_{\phi \tau}).
  \end{equation}
 On the other hand, the excitation path, obtained by inserting $r(s)$ and $g(s)$ into into \eref{eq:G_def}, is given by
 \begin{equation}
 k_{\uparrow}(x,\tau) = \mathcal{C}(\boldsymbol \rho^\star_{\phi \tau}) \frac{\int_0^1dx'  \mathcal{A}(\boldsymbol \rho^\star_{\phi \tau})}{\int_0^1dx'  \mathcal{C}(\boldsymbol \rho^\star_{\phi \tau})}-\mathcal{A}(\boldsymbol \rho^\star_{\phi \tau}) \frac{\int_0^1dx'  \mathcal{C}(\boldsymbol \rho^\star_{\phi \tau})}{\int_0^1dx'  \mathcal{A}(\boldsymbol \rho^\star_{\phi \tau})},
 \end{equation}
 and therefore in general $ k_{\uparrow}(x,\tau) \neq - k_{\downarrow}(x,-\tau) $. Of course the integral over the local current does obeys time-reversal symmetry,
 $ \int_0^1 dx k_{\uparrow}(x,\tau) = - \int_0^1 dx  k_{\downarrow}(x,-\tau)$, leading to the symmetry observed in the dynamics of $r$.
 The breaking of the time-reversal symmetry of $k(x,\tau)$ occurs in the ordered phase of the ABC model, where
   $\rho^\star$ is not homogenous \footnote{The time-reversal symmetry of $k(x,\tau)$ is not broken in the equal-densities
 ABC, where $\mathcal{C}(\boldsymbol \rho^\star)$ and $\mathcal{A}(\boldsymbol \rho^\star)$, which are chosen such that they maintain detailed-balance with respect to the conserving measure, are homogenous in space.}. Hence,
  the non-equilibrium nature of the system is maintained even in the $\phi\to0$ limit.

\section{Conclusions}
\label{sec:conc}

The phase diagram of the ABC model with slow nonconserving processes
is analyzed in the case where the overall densities of the three species are not
equal and the model is thus out of equilibrium. The phase diagram exhibits
features similar to those characterizing equilibrium systems with
long-range interactions such as ensembles inequivalence and negative
compressibility. It also exhibits features which are specific to nonequilibrium
systems, such as steady states with moving density profiles.

In the first part of the paper we focused on the case
where the non-conserving processes occur at a vanishingly slow
rate, $p\sim L^{-\gamma}, \gamma>2$, compared with rates of the conserving dynamics,
$\tau^{-1}\sim L^{-2}$. In
this limit, the dynamics of
the overall density, r, is shown to obey detailed balance although the
model itself does not. This allows one to derive the large deviations
function of r and draw from it the phase diagram of the
non-conserving model using equilibrium concepts such as
the definition of a chemical in the conserving model and
the calculation of the first order transition points using
the Maxwell construction. As is typical in equilibrium
systems with long-range interactions the conserving and non-conserving
models are found to exhibit the same critical lines but different
first order lines.

In the second part of the paper we studied the limit
where the conserving and non-conserving processes occur
at comparable time-scales, defined as $p=\phi L^2$. In
this case the models are found to differ not only in their first order
lines but also in their second order ones. In addition,
it is found that in the ordered phase the non-conserving
model exhibits a moving density profile with non-vanishing velocity in the
thermodynamic limit.

Using macroscopic fluctuation theory, we derived an
expansion in small $\phi$ of the instanton path leading to a
rare value of r. The leading order term of the action over
this instanton is identical to the expression of $\mathcal{G}(r)$ obtained
for $\gamma > 2$. This implies that $\gamma > 2$ is indeed the
limit where the conserving and non-conserving timescale
are well-separated. An interesting result of this
expansion is that all the correction terms in $\phi$ vanish in
the homogenous phase, implying that the expression
of $\mathcal{G}(r)$ obtained for $\gamma > 2$ is correct in the homogenous
phase for $\gamma = 2$ as well. This conclusion is valid for a
large class of driven-diffusive models that exhibit a homogenous
steady-state profile.

\begin{acknowledgements}
We thank A. Bar, L. Bertini, M. R. Evans, O. Hirschberg and J. L. Lebowitz for helpful discussions.
 The support of the Israel Science Foundation (ISF)  and of the Minerva Foundation with funding from the Federal German Ministry for Education and Research is
is gratefully acknowledged. The authors would like to thank the kind hospitality of the Korean Institute of Advanced Studies (KIAS) where
part of this work was conducted.
\end{acknowledgements}

\appendix

\section{Critical expansion of the ABC model with non-equal densities and vanishingly slow non-conserving dynamics ($p\ll L^{-2}$)}
\label{sec:high_order}

In this appendix we carry out the critical expansion of the conserving and the non-conserving ABC models, in the limit $p\sim L^{-\gamma}$ with $\gamma>2$, up to the forth order.
The expansion for the sixth order terms in both cases have been derived but will not be displayed due to their length.
They are used to determine the nature of the multi-critical point which connects the second and first order transition manifolds.

The starting point of the expansion is the dynamics of the density profile, given in the conserving model by
\begin{equation}
\label{eq:meanfieldA_app}
\partial_{\tau} \rho_\alpha = \beta \partial_x\left[\rho_\alpha\left(\rho_{\alpha+1}-\rho_{\alpha+2}\right)\right]+ \partial_x^2\rho_\alpha.
\end{equation}
The non-conserving equation is obtained by adding a term of the form $L^2 p \left(\rho_0^3-e^{-3\beta\mu}\rho_A\rho_B\rho_C\right)$ to the right hand
side of \eref{eq:meanfieldA_app}.
As discussed in \sref{sec:expand}, in both the conserving and the non-conserving models the density profile can be expanded near the critical line in terms of the following Fourier modes:
\begin{eqnarray}
\label{eq:rho_expand_app}
\fl \left(\begin{array}{c}
\rho_{A}\\
\rho_{B}\\
\rho_{C}
\end{array}\right)&=&\sum_{m=-\infty}^{\infty}e^{2\pi imx}\Big[\frac{a_{m}(\tau)}{\sqrt{3}}\left(\begin{array}{c}
1\\
e^{-2\pi i/3}\\
e^{2\pi i/3}
\end{array}\right) \\
&&+\frac{a_{-m}^{\star}(\tau)}{\sqrt{3}}\left(\begin{array}{c}
1\\
e^{2\pi i/3}\\
e^{-2\pi i/3}
\end{array}\right)+\frac{b_{m}(\tau)}{3}\left(\begin{array}{c}
1\\
1\\
1
\end{array}\right)\Big]. \nonumber
\end{eqnarray}

We first analyze the expansion of the conserving ABC model. Inserting \eref{eq:rho_expand_app} into \eref{eq:meanfieldA_app} yields a set of decoupled equations of motion, given by
\begin{eqnarray}
\label{eq:dam_app}
\partial_{\tau}a_{m}&=&-\frac{2\pi m}{3}\big[\left(6\pi m-\sqrt{3}\beta r\right)a_{m} \\
&&\qquad +3\beta\sum_{m'=-\infty}^{\infty}a_{-m-m'}^{\star}a_{m'}^{\star}\big],\nonumber \\
\label{eq:dbm_app}
\partial_{\tau} b_m &=& - 4\pi ^2 m^2 b_m.
\end{eqnarray}
The latter equation implies that in the steady-state $b_m=0$ for $m\neq 0$ and $b_0=r$.
This corresponds to a flat profile of the vacancies which is a result of their unbiased dynamics.
It is useful to rewrite \eref{eq:dam_app} in the following vector form:
\begin{eqnarray}
\label{eq:dam1_app}
\left(\begin{array}{c}
\partial_{\tau}a_{m}\\
\partial_{\tau}a_{-m}^{\star}
\end{array}\right)&=&-\frac{2\pi m}{3}\Big[A_{m}\left(\begin{array}{c}
a_{m}\\
a_{-m}^{\star}
\end{array}\right)  \\
&&\qquad +3\beta\sum_{m'=-\infty}^{\infty}\left(\begin{array}{c}
a_{-m-m'}^{\star}a_{m'}^{\star}\\
-a_{m-m'}a_{m'}
\end{array}\right)\Big], \nonumber
\end{eqnarray}
where
\begin{equation}
A_m=\left(\begin{array}{cc}
6\pi m-\sqrt{3}\beta r & 6\beta a_{0}^{\star}\\
-6\beta a_{0} & 6\pi m+\sqrt{3}\beta r
\end{array}\right).
\end{equation}
%It is worth noting that the vanishing of the off-diagonal terms in $A_m$ in the equal-densities case, $a_0=0$, greatly simplifies the analysis below, presented for a general $a_0$.
The eigenvalues of $-\frac{2\pi m}{3} A_m$ are
\begin{equation}
\epsilon_{\pm}^{(m)}(\beta,r,a_0)= \frac{2\pi m}{3} \left( - 6 m \pi \pm  \beta \sqrt{3 \left( r^2-12|a_{0}|^{2}\right)}\right),
\end{equation}
and the corresponding left and right eigenvectors are independent of $m$ and given by
\begin{eqnarray}
  v_{\pm}^{R}&=&\frac{1}{C_{\pm}}\left(\begin{array}{c}
r\pm\sqrt{r^{2}-12|a_{0}|^{2}}\\
2\sqrt{3}a_{0}
\end{array}\right), \\
v_{\pm}^{L}&=&\frac{1}{C_{\pm}}\left(\begin{array}{c}
-r\mp\sqrt{r^{2}-12|a_{0}|^{2}}\\
2\sqrt{3}a_{0}^{\star}
\end{array}\right),
\end{eqnarray}
where $C_{\pm}=2\sqrt{r^{2}-12|a_{0}|^{2}}(r\mp\sqrt{r^{2}-12|a_{0}|^{2}})$ is a normalization factor.
When the highest eigenvalue vanishes, $\epsilon_{+}^{(1)}=0$, the corresponding Fourier mode becomes unstable, leading to an ordered profile.
The critical line is defined as $\epsilon_{+}^{(1)}(\beta_c,r,a_0)=0$, where $\beta_c$ is given in \eref{eq:bc_cons_non-equal}.

In order to determine whether the conserving model exhibits a second order phase transition on the critical line one has to check whether for $\beta=\beta_c(1+\delta)$
the model exhibits an order phase whose amplitude vanishes with $\delta$. To this end, it is useful to denote the first unstable Fourier mode as
\begin{equation}
\label{eq:app_phi_def}
\left(\begin{array}{c}
a_{1}\\
a_{-1}^{\star}
\end{array}\right) = \varphi  v_+^R +    O(\varphi^2)  v_+^L,
\end{equation}
 where $\varphi$ is given by  up to a constant by some positive power of $\delta$, which will be determined below. Assuming a steady state, $\partial_t a_m=0$, and inserting \eref{eq:app_phi_def}  into \eref{eq:dam1_app} for $m=2$, yields to leading order
\begin{equation}
\label{eq:dam2_app}
\left(\begin{array}{c}
a_{2}\\
a_{-2}^{\star}
\end{array}\right)=3\beta_{c}\varphi^{2}A_{2}^{-1}\left(\begin{array}{c}
-(v_{+,2}^{R})^{2}\\
(v_{+,1}^{R})^{2}
\end{array}\right)+O(\varphi^{4}),
\end{equation}
where $v^R_{+,1}$ and $v^R_{+,2}$ denote the two elements of $v^R_+$.
Inserting this back into \eref{eq:dam1_app} for $m=1$, yields the forth order term in its expansion in powers of $\varphi$, given by
\begin{equation}
0= 4 \pi ^2 \delta \varphi +   G_{4}^{c}(r,a_{0}) \varphi |\varphi|^2 + O(\varphi|\varphi|^4),
\end{equation}
where the $4\pi^2$ factor is due to fact that $\epsilon_{+}^{(1)}(\beta_c(1+\delta),r,a_0)=4\pi^2 \delta$ and
\begin{eqnarray}
 G_{4}^{c}(r,a_{0})& =&  \frac {1}{ \varphi |\varphi|^2} 6\beta_c v_{+}^{L}\cdot\left(\begin{array}{c}
a_{-2}^{\star}a_{1}^{\star}\\
-a_{2}a_{-1}
\end{array}\right)  \\
& =&   \frac{32 \pi ^2 \left(r^{2}+r\sqrt{r^{2}-12|a_{0}|^{2}}
-6|a_{0}|^{2}\right)}{\left(r^{2}-12|a_{0}|^{2}\right)^{2}\left(r+\sqrt{r^{2}-12|a_{0}|^{2}}\right)} \nonumber \\
&&  \times \left(r^{3}-18r|a_{0}|^{2}-12\sqrt{3}|a_{0}|^{3}\cos(3\theta)\right), \nonumber
\end{eqnarray}
with $\theta \equiv \mathrm{arg}(a_0)$.

\begin{figure}[t]
\noindent
\begin{centering}\includegraphics[scale=0.6]{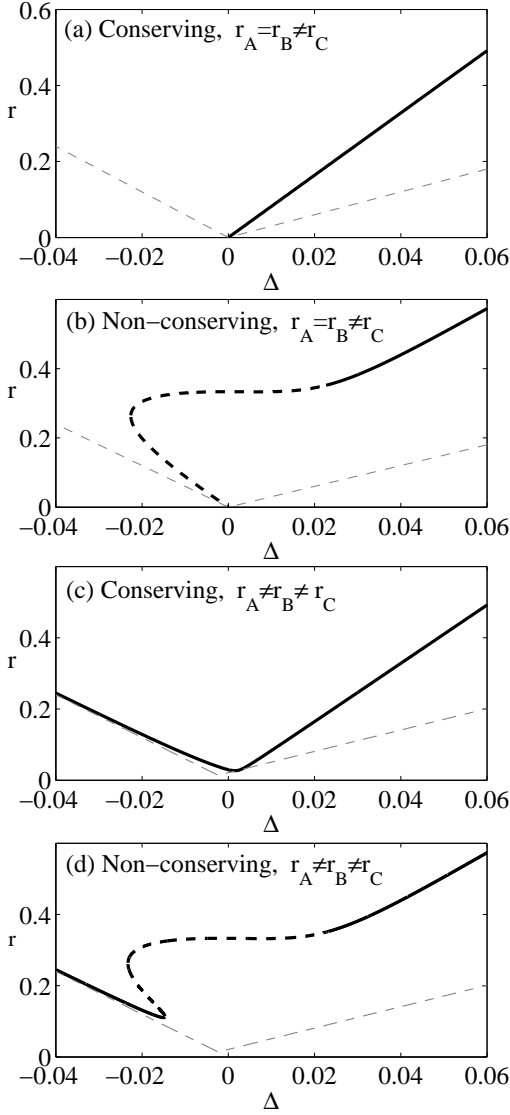}\par\end{centering}
\caption{  Tricritical lines in the conserving and the non-conserving models. The dashed
and solid thick lines represent points where $G_6 < 0$ and $G_6 > 0$, respectively. In the latter case the tricritical points are preempted by a critical-end-points.
The thin dashed gray lines denote the smallest possible values of $r$. The case of two equal
densities, $r_A = r_B = r/3-\Delta$ and $r_C = r/3+2\Delta$ or equivalently $a_0=\Delta \sqrt{3} e^{i\pi /3}$ , is shown in (a) and (b).
Figures (c) and (d) show the fully non-equal-densities case, where $r_A = r/3-\Delta+0.007$, $r_B = r/3-\Delta+0.007$ and $r_C = r/3+2\Delta$.
\label{fig:TCL}}
\end{figure}

For $G_{4}^{c}(r,a_{0})>0$ one finds that $\varphi \sim \delta^{1/2}$, which implies that the model exhibits a second order phase transition on the critical line.
On the other hand, when $G_{4}^{c}(r,a_{0})<0$ the critical line is preempted by a first order phase transition. On the line where $G_{4}^{c}(r,a_{0})=0$,
plotted in figures \ref{fig:TCL}a and \ref{fig:TCL}c, a further
analysis of the expansion shows that $G_{6}^{c}(r,a_{0})>0$, which implies that the second order
and first order transition lines join in the conserving model at a tricrtical point.

In the non-conserving model the additional term in the dynamics of $\boldsymbol \rho(x,\tau)$ affects directly only the equation of $b_m$. Since this term scales as $pL^2$, it is negligible in comparison to the diffusive term in \eref{eq:dbm_app}, $- 4\pi ^2 m^2 b_m$, for $m\neq 0$. For $m=0$ the non-conserving term yields the following equation:
\begin{eqnarray}
\label{eq:db0_app}
\fl \partial_{\tau} b_0 &= & 3 p L^2 \int_0^1 dx \left ( \rho_0^3 -e^{-3\beta \mu}\rho_A\rho_B\rho_C\right)  \\
\fl &=&  3 p L^2 \Big\{ (1-b_0)^3-e^{-3\beta \mu} \big[ \frac{b_0^3}{27}-\frac{b_0}{3}  \sum_{m_1} a_{m_1}a^\star_{m_1} \nonumber  \\
&&+\frac{1}{3\sqrt{3}} \sum_{m_1,m_2} (a_{m_1}a_{m_2}a_{-m_1-m_2} + c.c.) \big] \Big\}. \nonumber
\end{eqnarray}
It is useful to consider the fluctuations in the overall density are denoted by $\delta r$, such that $b_0=\rs+\delta r$ and $\rs$ is given by \eref{eq:rmu}.
The parameter $\delta r$ enters the equation of $a_m$ in the following form:
\begin{eqnarray}
\label{eq:dam1_appA}
\partial_{\tau}a_{m}&=&-\frac{2\pi m}{3}\Big[\left(6\pi m-\sqrt{3}\beta \rs\right)a_{m} -\sqrt{3} \beta \delta r a_m \nonumber \\
&&+3\beta\sum_{m'=-\infty}^{\infty}a_{-m-m'}^{\star}a_{m'}^{\star}\Big].
\end{eqnarray}
In the steady-state, \eref{eq:db0_app} is shown below to yield to leading order  $\delta r = \mathcal{O}(|a_m|^2, a_m a_{-m})$.
As a result, the fluctuation in the density, $\delta r$, do not
affect the linear stability of $a_m$ and thus the critical line of the non-conserving model is identical to that of the conserving model.

The type of transition observed on the critical line is determined by the higher order terms in the expansion.
For convenience the expansion is expressed below in terms of the steady-state density, $\rs$, instead of $\mu$. Inserting the definition for $\varphi$ in \eref{eq:app_phi_def}
into \eref{eq:db0_app} yields to leading order
\begin{equation}
\label{eq:deltar_app}
\delta r=\frac{6(1-\rs)\left(\rss-12|a_{0}|^{2}\right)\left(\rs+\sqrt{\rss-12|a_{0}|^{2}}\right)}{\rss-3(1+2\rs)|a_{0}|^{2}}|\varphi|^{2}.
\end{equation}
The $\mathcal{O}(\varphi^4)$ term in the expansion of \eref{eq:dam1_appA} is then obtained by inserting the
leading order term of $a_2$ in \eref{eq:dam2_app} and of $\delta r$ in \eref{eq:deltar_app} into \eref{eq:dam1_appA}, yielding
\begin{equation}
0= 4 \pi ^2 \delta \varphi + G_{4}^{nc}(\rs,a_{0}) \varphi |\varphi|^2 + O(\varphi|\varphi|^4),
\end{equation}
where
\begin{eqnarray}
  && G_{4}^{nc}(r,a_{0})  = G_{4}^{c}(r,a_{0}) -\frac {1}{ \varphi |\varphi|^2} \sqrt{3} \beta \delta r v_{+}^{L}\cdot\left(\begin{array}{c}
 a_1\\
- a_{-1}^\star
\end{array}\right) \nonumber \\
 && =  G_{4}^{c}(r,a_{0}) - \frac{24 \pi^2 r(1-r)\left(r+\sqrt{r^{2}-12|a_{0}|^{2}}\right)}{r^{2}-3(1+2r)|a_{0}|^{2}}.
\end{eqnarray}

In the equal-densities case the tricritical point is found at $\rs=1/3$, $G_{4}^{nc}(1/3,0)=0$, as previously found in \cite{Lederhendler2010b}.
 For $a_0\neq 0$, by continuing the expansion described above it can be shown that $G_{6}^{nc}(r,0)<0$ on the tricritical line for values of $|a_0|$ that are relatively small.
 In this case, the tricritical point is preempted by a critical-end-point, which connects the first and second order transition line, as demonstrated in \fref{fig:MuT_non-equal}b.
Figures \ref{fig:TCL}b and \ref{fig:TCL}d show the tricritical line of the non-conserving model, defined by $\beta=\beta_c$ and $G_{4}^{nc}(r,a_0)=0$, in a specific
 cross-section of parameter-space.
The dashed and solid lines denote regions where $G_{6}^{nc}(r,a_0)<0$  and $G_{6}^{nc}(r,a_0)>0$ on the tricritical line, respectively.

\section{Derivation of Hamilton's equations for the non-conserving ABC model with $p=\phi L^{-2}$}
\label{sec:hamilton}
In \sref{sec:fast_fol} the non-conserving ABC model is analyzed in the limit where $p=\phi L^{-2}$ using the macroscopic fluctuation theorem.
 In this appendix we derive a set of equations
for the instanton trajectory that minimizes the action of the model. The trajectory is defined in terms of fields that can be regarded as the momenta conjugate to the conserving
and non-conserving particles currents. In this respect, the equations derived below play the role of Hamilton's equations.

In the Lagrangian formulation presented in \sref{sec:fast_fol},  the instanton path is defined by the currents $\mathbf{j}(x,\tau)$ and $k(x,\tau)$ and the corresponding action is given by
\begin{eqnarray}
\mathcal{I} &=& \int_{0}^{{T}}d\tau\int_{0}^{1}dx\Big[\mathcal{L}_{c}(\boldsymbol{\rho}(x,\tau),\mathbf{j}(x,\tau)) \\
&&+\phi\mathcal{L}_{nc}(\boldsymbol{\rho}(x,\tau),k(x,\tau))\Big]. \nonumber
\end{eqnarray}
Here the conserving and the non-conserving Lagrangian densities are given by
\begin{equation}
\mathcal{L}_{c}(\mathbf{j},\boldsymbol{\rho})=\left[\mathbf{j}+\partial_{x}\boldsymbol{\rho}
+\boldsymbol \chi(\boldsymbol{\rho})\right]\Sigma^{-1}(\boldsymbol{\rho})\left[\mathbf{j}+\partial_{x}\boldsymbol{\rho}+{\boldsymbol \chi}(\boldsymbol{\rho})\right],
\end{equation}
and
\begin{eqnarray}
\mathcal{L}_{nc}(\boldsymbol{\rho},k)&=&\mathcal{C}(\boldsymbol{\rho})+\mathcal{A}(\boldsymbol{\rho})
-\sqrt{k^{2}+4\mathcal{A}(\boldsymbol{\rho})\mathcal{C}(\boldsymbol{\rho})} \nonumber \\
&&+k\ln(\frac{\sqrt{k^{2}+4\mathcal{A}(\boldsymbol{\rho})\mathcal{C}(\boldsymbol{\rho})}+k}{2\mathcal{C}(\boldsymbol{\rho})}),
\end{eqnarray}
respectively. The density profile of the particles along this path is given by the continuity equation:
\begin{equation}
\label{eq:app_continuity}
\partial_\tau \boldsymbol{\rho}(x,\tau) = -\partial_x \mathbf{j}(x,\tau) + \phi {\bf u} k(x,\tau),
\end{equation}
where ${\bf u} =(1,1,1)^T$.
In the derivation below we consider a general form of  ${\boldsymbol \chi}(\boldsymbol{\rho}), \mathcal{A}(\boldsymbol{\rho})$ and $\mathcal{C}(\boldsymbol{\rho})$.
 The specific expression of these functionals are discussed in \sref{sec:fast_fol}.

As mentioned in \sref{sec:fast_fol}, it is convenient to express the instanton path using the fields ${\bf H}(x,\tau)$ and $G(x,\tau)$, defined by the following equations:
\begin{eqnarray}
\label{eq:app_HG_deff}
\Sigma (\boldsymbol \rho) \partial_x {\bf H} = {\bf j} + \partial_x{\boldsymbol \rho} + {\boldsymbol \chi}(\boldsymbol \rho),\\
k=\mathcal{C}(\boldsymbol \rho)e^{G} - \mathcal{A}(\boldsymbol \rho)e^{-G}.
\end{eqnarray}
In terms of these fields the Lagrangian densities are given by
\begin{eqnarray}
\label{eq:app_Hamiltonian_action}
 \mathcal{L}_c(\boldsymbol \rho, H)&=&\frac{1}{2}(\partial_x {\bf H})^T\Sigma(\boldsymbol\rho) \, \partial_x {\bf H}, \\
 \mathcal{L}_{nc}(\boldsymbol \rho, H) &=& \mathcal{C}(\boldsymbol \rho)(1-e^{G}+G e^{G})  \\
  && + \mathcal{A}(\boldsymbol \rho)(1-e^{-G}-G e^{-G}),  \nonumber
\end{eqnarray}
and the continuity equations are given by
\begin{equation}
\label{eq:app_continuity1}
\partial_{\tau}\boldsymbol{\rho}=-\partial_{x}[\Sigma(\boldsymbol{\rho})\partial_{x}{\bf H}-\partial_{x}\boldsymbol{\rho}-\boldsymbol{\chi}(\boldsymbol{\rho})]+\phi{\bf u}[C(\boldsymbol{\rho})e^{G}-A(\boldsymbol{\rho})e^{-G}].
\end{equation}
In \eqs(\ref{eq:app_HG_deff})-(\ref{eq:app_continuity1}) and in the derivation below the
$(x,\tau)$-dependence is omitted in some cases in order to keep the notation compact.

We denote by $\bar {\bf H}$ and $\bar{G}$ the path that minimizes $\mathcal{I}$ and consider small deviations around this path defined as,
\begin{eqnarray}
\label{eq:app_HG_expand}
{\bf H}(x,\tau)&=&\bar {\bf H}(x,\tau) +{\boldsymbol \delta}_{H}(x,\tau), \\
 G(x,\tau)&=& \bar{G}(x,\tau)+\delta_G (x,\tau). \nonumber
\end{eqnarray}
Similarly using the continuity equations one can define the fluctuation of the density profile around the optimal path, as
\begin{eqnarray}
\label{eq:app_rho_expand}
{\boldsymbol \rho}(x,\tau)&=&\bar{\boldsymbol \rho}(x,\tau)+{\boldsymbol \delta}_{\rho}(x,\tau),
\end{eqnarray}
where $\bar{\boldsymbol \rho}$ is obtained by inserting $\bar{\bf H}$ and $\bar{G}$ into \eref{eq:app_continuity1}.
The bar notation here should not be confused with that used in \sref{sec:fast_fol} to denote the density profile at the end of the instanton path.

Below we derive the first order correction to the action due to ${\boldsymbol \delta}_{H}$ and $\delta_G$, defined as
\begin{eqnarray}
\label{eq:app_I_expand}
\mathcal{I} & =&\int_{0}^{{T}}d\tau\int_{0}^{1}dx[\mathcal{L}_{c}(\boldsymbol{\bar{\boldsymbol{\rho}}},\bar{{\bf H}}) \\
 && \qquad +\mathcal{L}_{nc}(\boldsymbol{\bar{\boldsymbol{\rho}}},\bar{G})+\delta\mathcal{L}^{(1)}+\mathcal{O}(\delta_{H,\alpha}^{2},\delta_{G}^{2})]. \nonumber
\end{eqnarray}
The optimal trajectory is defined by the equation $\delta\mathcal{L}^{(1)}=0$.
As shown below, it is in fact more convenient to express this equation in terms of the fields ${\boldsymbol \delta}_{\rho}$ and $\delta_G$.

The simpler part in the first order expansion of $\mathcal{I}$ is the expansion of the non-conserving Lagrangian density in terms of $\delta_G$ and $\boldsymbol \delta_\rho$.
Inserting \eref{eq:app_HG_expand} and \eref{eq:app_rho_expand} into the form of $\mathcal{L}_{nc}$ yields the following expansion:
\begin{eqnarray}
\label{eq:app_Lnc_expand}
\fl && \delta\mathcal{L}_{nc}(\boldsymbol{\rho},G)=\mathcal{L}_{nc}(\boldsymbol{\boldsymbol{\rho}},G) -\mathcal{L}_{nc}(\boldsymbol{\bar{\boldsymbol{\rho}}},\bar{G}) \nonumber \\
\fl && =\sum_{\alpha}\delta_{\rho,\alpha}[\partial_{\rho_{\alpha}}\mathcal{C}(\boldsymbol{\bar{\boldsymbol{\rho}}})(1-e^{\bar{G}}+\bar{G}e^{\bar{G}})  \\
&& \qquad +\partial_{\rho_{\alpha}}\mathcal{A}(\boldsymbol{\bar{\boldsymbol{\rho}}})(1-e^{-\bar{G}}-\bar{G}e^{-\bar{G}})] \nonumber \\
 && \qquad+
\phi\delta_{G}[\mathcal{C}(\boldsymbol{\bar{\boldsymbol{\rho}}})\bar{G}e^{\bar{G}}+\mathcal{A}(\boldsymbol{\bar{\boldsymbol{\rho}}})\bar{G}e^{-\bar{G}}]
+\mathcal{O}(\delta_{\rho,\alpha}^{2},\delta_{G}^{2}).\nonumber
\end{eqnarray}

The conserving Lagrangian density should be expanded first in terms of $\boldsymbol \delta_H$ and $\boldsymbol \delta_\rho$, yielding
\begin{eqnarray}
\label{eq:app_Lc_expand}
&&\delta\mathcal{L}_{c}(\boldsymbol{\rho},{\bf H})=\frac{1}{2}(\partial_{x}{\bf H})^{T}\Sigma(\boldsymbol{\rho})\,\partial_{x}{\bf H}-\mathcal{L}_{c}(\boldsymbol{\bar{\boldsymbol{\rho}}},\bar{{\bf H}})  \\
&& = \partial_{x}\bar{{\bf H}}^{T}\Sigma(\bar{\boldsymbol{\rho}})\partial_{x}\boldsymbol{\delta}_{H}+\frac{1}{2}\sum_{\alpha}\delta_{\rho,\alpha}(\partial_{x}\bar{{\bf H}})^{T}\partial_{\rho_{\alpha}}\Sigma(\bar{\boldsymbol{\rho}})\partial_{x}\bar{{\bf H}}. \nonumber
\end{eqnarray}
The first term in the line above can be further simplified by employing integration by parts over $x$ of this term, yielding
\begin{eqnarray}
\label{eq:app_K1}
\fl K &\equiv& \int_{0}^{{T}}d\tau  \int_{0}^{1}dx \partial_{x}\bar{{\bf H}}^{T}\Sigma(\bar{\boldsymbol{\rho}})\partial_{x}\boldsymbol{\delta}_{H} \\
 &=&
-\int_{0}^{{T}} d\tau \int_{0}^{1}dx \bar{{\bf H}}^{T}\partial_{x}[\Sigma(\bar{\boldsymbol{\rho}})\partial_{x}\boldsymbol{\delta}_{H}]. \nonumber
\end{eqnarray}
Here and below the boundary terms of the integration by parts over $x$ vanish due to the periodic boundary condition.
As mentioned above, it is convenient to express  the terms in $\delta\mathcal{L}^{(1)}$ that involve ${\boldsymbol \delta}_{H}$  in terms of the fields ${\boldsymbol \delta}_{\rho}$ and $\delta_G$.
This can be done by inserting \eqs(\ref{eq:app_HG_expand}) and (\ref{eq:app_rho_expand}) into the continuity equation (\ref{eq:app_continuity1}), yielding
\begin{eqnarray}
\fl&& \partial_{x}[\Sigma(\bar{\boldsymbol{\rho}})\partial_{x}\delta_{H}]_{\alpha}=-\partial_{\tau}\delta_{\rho,\alpha}+\partial_{x}^{2}\delta_{\rho,\alpha} \\
&&\qquad +\sum_{\alpha'}\partial_{x}[\delta_{\rho,\alpha'}(\partial_{\rho_{\alpha'}}\Sigma(\bar{\boldsymbol{\rho}})\partial_{x}\bar{{\bf \ H}})_{\alpha}
+\delta_{\rho,\alpha'}\partial_{\rho_{\alpha'}}\chi_{\alpha}(\bar{\boldsymbol{\rho}})]  \nonumber \\
&&\qquad +\phi\delta_{G}[C(\bar{\boldsymbol{\rho}})e^{\bar{G}}+A(\bar{\boldsymbol{\rho}})e^{-\bar{G}}]. \nonumber
\end{eqnarray}
Inserting this expression into \eref{eq:app_K1} yields an integral which involves derivatives of $\boldsymbol \delta_\rho$ and $\delta_G$.
We wish, however, to obtain an expression of $\delta\mathcal{L}^{(1)}$ which involves only local terms of the variational fields. This can be done
using the proper integrations by parts over $x$ and $\tau$, which yield
\begin{eqnarray}
\label{eq:app_K2}
 K &=&\int_{0}^{{T}}d\tau\int_{0}^{1}dx\sum_{\alpha}\{\delta_{\rho,\alpha}[-\partial_{\tau}\bar{H}_{\alpha} \\
&&-\partial_{x}\bar{{\bf H}}^{T}\partial_{\rho_{\alpha}}\Sigma(\bar{\boldsymbol{\rho}})\partial_{x}\bar{{\bf H}}-\partial_{x}^{2}\bar{H}_{\rho,\alpha}-\partial_{x}\bar{{\bf H}}^{T}\partial_{\rho_{\alpha}}\boldsymbol{\chi}(\boldsymbol{\bar{\boldsymbol{\rho}}})] \nonumber \\
&&-\phi\delta_{G}\bar{H}_{\alpha}[C(\boldsymbol{\bar{\boldsymbol{\rho}}})e^{\bar{G}}+A(\bar{\boldsymbol{\rho}})e^{-\bar{G}}]\}+\mathcal{O}(\delta_{\rho,\alpha}^{2},\delta_{G}^{2}). \nonumber
\end{eqnarray}

Finally, by inserting \eqs(\ref{eq:app_Lnc_expand})-(\ref{eq:app_K2}) into \eref{eq:app_I_expand} one obtains the following
expression for the first order term in the expansion of the action:
\begin{eqnarray}
 \delta\mathcal{L}^{(1)}&=&\sum_{\alpha}\delta_{\rho,\alpha}[-\partial_{\tau}\bar{H}_{\alpha}-\frac{1}{2}\partial_{x}\bar{{\bf H}}^{T}\partial_{\rho_{\alpha}}\Sigma(\bar{\boldsymbol{\rho}})\partial_{x}\bar{{\bf H}} \nonumber \\
 && -\partial_{x}^{2}\bar{H}_{\rho,\alpha}-\partial_{x}\bar{{\bf H}}^{T}\partial_{\rho_{\alpha}}\boldsymbol{\chi}(\boldsymbol{\bar{\boldsymbol{\rho}}})] \\
 &&-\phi\delta_{G}(\bar{G}-\sum_{\alpha}\bar{H}_{\alpha})
[C(\boldsymbol{\bar{\boldsymbol{\rho}}})e^{\bar{G}}+A(\bar{\boldsymbol{\rho}})e^{-\bar{G}}]. \nonumber
\end{eqnarray}
The solution for $\delta\mathcal{L}^{(1)}=0$ corresponds to the extremum of the action, which is therefore given by
\begin{eqnarray}
 \label{eq:app_Hamil_G}
  \bar{G}&=&  \bar{H}_A+\bar{H}_B+\bar{H}_C\\
 \label{eq:app_Hamil_H}
 \partial_\tau \bar{H}_\alpha & = &  - \partial_x^2 \bar{{\bf H}}^T  + \partial_{\rho_\alpha} \boldsymbol \chi (\bar{\boldsymbol \rho})\partial_x \bar{H}_\alpha -\frac{1}{2}(\partial_x {\bar{\bf H}})^T\partial_{\rho_\alpha}\Sigma (\bar{\boldsymbol \rho}) \, \partial_x \bar{{\bf H}} \nonumber \\
& &+  \phi [\partial_{\rho_\alpha} \mathcal{C}(\bar{\boldsymbol \rho})(1-e^{\bar{G}}) + \partial_{\rho_\alpha} \mathcal{A}(\bar{\boldsymbol \rho})(1-e^{-\bar{G}})].
\end{eqnarray}
The instanton path is fully described by the two equations above and the definition of $\bar {\boldsymbol \rho}$, given by
\begin{equation}
\partial_{\tau}\bar{\boldsymbol{\rho}}=-\partial_{x}[\Sigma(\bar{\boldsymbol{\rho}})\partial_{x}\bar{\bf H}-\partial_{x}\bar{\boldsymbol{\rho}}-\boldsymbol{\chi}(\bar{\boldsymbol{\rho}})]+\phi{\bf u}[\mathcal{C}(\bar{\boldsymbol{\rho}})e^{\bar{G}}-\mathcal{A}(\bar{\boldsymbol{\rho}})e^{-\bar{G}}].
\end{equation}
In \sref{sec:fast_fol} the bar notation is omitted from the above result in order to simplify the notation.

\bibliographystyle{apsrev4-1}
\bibliography{ABCModel}

\end{document}